\documentclass[jgr,times]{agutex2arxiv}
\usepackage{siunitx}
\usepackage{xcolor,amsmath,amssymb}
\usepackage{graphicx, mathptmx}
\pdfoutput=1

\authorrunninghead{DAVID ET AL.}
\titlerunninghead{}
\authoraddr{E. C. David, e.david@ucl.ac.uk}

\begin{document}

\title{Absence of Stress-induced Anisotropy during Brittle Deformation in Antigorite Serpentinite}

\authors{Emmanuel C. David\altaffilmark{1},
  Nicolas Brantut\altaffilmark{1}, Lars N. Hansen\altaffilmark{2},
  and Thomas M. Mitchell\altaffilmark{1}}

\altaffiltext{1}{Department of Earth Sciences,
  University College London, London, United Kingdom.}
\altaffiltext{2}{Department of Earth Sciences, University of Oxford,
  Oxford, United Kingdom.}

\begin{abstract}
  Knowledge of the seismological signature of serpentinites during deformation is fundamental for interpreting seismic observations in subduction zones, but this has yet to be experimentally constrained. We measured compressional and shear wave velocities during brittle deformation in polycrystalline antigorite, at room temperature and varying confining pressures up to 150~MPa. Ultrasonic velocity measurements, at varying directions to the compression axis, were combined with mechanical measurements of axial and volumetric strain, during direct loading and cyclic loading triaxial deformation tests. An additional deformation experiment was conducted on a specimen of Westerly granite for comparison. At all confining pressures, brittle deformation in antigorite is associated with a spectacular absence of stress-induced anisotropy and with no noticeable dependence of wave velocities on axial compressive stress, prior to rock failure. The strength of antigorite samples is comparable to that of granite, but the mechanical behaviour is elastic up to high stress ($\gtrsim$80$\%$ of rock strength) and non-dilatant. Microcracking is only observed in antigorite specimens taken to failure and not in those loaded even at 90--95$\%$ of their compressive strength. Microcrack damage is extremely localised near the fault and consists of shear microcracks that form exclusively along the cleavage plane of antigorite crystals. Our observations demonstrate that brittle deformation in antigorite occurs entirely by ``mode II'' shear microcracking. This is all the more remarkable than the preexisting microcrack population in antigorite is comparable to that in granite. The mechanical behaviour and seismic signature of antigorite brittle deformation thus appears to be unique within crystalline rocks. 
\end{abstract}


\begin{article}

  \section{Introduction}
  \label{sec:intro}

  Serpentinites form by hydrothermal alteration of ultramafic rocks from the oceanic lithosphere, and are commonly found in and around mid-ocean ridges, transform faults, obducted ophiolites, and in the subducting slabs and the overriding mantle wedge within subduction zones. As such, they play a major role in controlling lithospheric strength (\textit{e.g.}, \citet{Escartinetal1997,HyndmanPeacock2003}), rheological behaviour (\textit{e.g.}, \citet{Hilairetetal2007,Amiguetetal2012,HirauchiKatayama2013,Auzendeetal2015}), frictional properties (\textit{e.g.}, \citet{Reinenetal1994,Mooreetal1997}), and mechanical anisotropy (\textit{e.g.}, \citet{Padronetal2012}) in subduction zones. Reviews of the occurence and tectonic significance of serpentinites in subduction zones were recently presented by \citet{Reynard2013} and \citet{Guillotetal2015}. Among the serpentine group of hydrous phyllosilicates (13$\%$ water in weight), formed of three polytypes -- lizardite, antigorite, and chrysotile, by decreasing order of abundance --  \emph{antigorite} is the mineral stable over the largest depth range \citep{UlmerTrommsdorff1995, Reynard2013}. Antigorite has an elongated and ``corrugated'' crystallographic structure due to the regular inversion of parallel tetrahedral and tri-octahedral sheets that regularly alternate in their ordering. These sheets define antigorite's basal plane \citep{Otten1993,WicksOHanley1988}, and confer to antigorite a strong anisotropy. For an extended summary of the physical properties of antigorite, see \citet{Reynard2013}.

  Triaxial deformation experiments demonstrate that antigorite serpentinites are brittle in a variety of hydrostatic pressure and temperature conditions at conventional laboratory strain rates (10$^{-3}$--10$^{-5}$~\si{\per\second}). At room temperature, the transition from localised to more distributed deformation is observed around 300--350~\si{\mega\pascal} confining pressure \citep{RaleighPaterson1968,MurrellIsmail1976,Escartinetal1997}. These studies also demonstrate that lizardite, chrysotile and assemblages of serpentine minerals are much weaker than antigorite. Brittle behaviour is also observed in antigorite above these pressures when temperature is increased, associated with the ``dehydration-embrittlement'' phenomenon \citep{RaleighPaterson1968,Jungetal2004}. However, recent deformation experiments show that brittle behaviour is also widely observed across the antigorite stability field even at elevated pressures and temperatures \citep{ChernakHirth2010,ProctorHirth2016,Gascetal2017}.

  Measurements of \emph{seismic properties} are crucial in order to better interpret geophysical data, \textit{i.e}, to detect and quantify the presence of serpentinites in subduction zones. Several studies have attempted to measure or constrain elastic wave velocities in antigorite serpentinites in a range of experimental conditions. At room temperature, the full elastic tensor of \emph{single-crystal antigorite} was measured using Brillouin spectroscopy under ambient conditions and high pressures up to \SI{9}{\giga\pascal} by \citet{Bezacieretal2010} and \citet{Bezacieretal2013}, respectively, from which isotropic aggregate properties were calculated using averaging methods. The elasticity of antigorite has also been calculated from equation-of-state measurements \citep{Hilairetetal2006} or computed using ab-initio calculations \citep{MookherjeeCapitani2011}, both demonstrating good agreement with experimental data. On \emph{polycrystalline antigorite serpentinites}, P- and S-wave ultrasonic velocities have been measured under hydrostatic compression by \citet{Birch1960}, \citet{Simmons1964} and \citet{Christensen1978}, up to \SI{1}{\giga\pascal}. While velocities are strongly dependent on serpentine mineralogy, and the presence of accessory minerals, antigorite is characterised by relatively high P- and S-wave velocities, which increase with pressure over several hundreds of \si{\mega\pascal} as a result of the gradual closure of microcracks. Attention to velocity anisotropy was given in subsequent measurements of \citet{Watanabeetal2007} and \citet{Kernetal1997} on strongly foliated antigorite serpentinites up to 200 and 600 \si{\mega\pascal}, respectively. The marked anisotropy measured on their samples (up to 30$\%$) results from combined effects of microcracks having preferred orientation and lattice-preferred orientation of major minerals \citep{Moralesetal2018}. Accordingly, anisotropy decreases with confining pressure as cracks gradually close up. More recently, \citet{Jietal2013} and \citet{Shaoetal2014} measured the pressure-dependence and anisotropy of P- and S-wave velocities on a large set of antigorite serpentinites under hydrostatic compression up to \SI{650}{\mega\pascal}. Their experimental results demonstrate that the intrinsic velocity anisotropy is mostly caused by lattice-preferred orientation of antigorite. The complex evolution of velocity anisotropy with hydrostatic pressure below the crack-closure pressure ($<$\SI{150}{\mega\pascal}) is rock-dependent and results from competing effects between microcrack and lattice-preferred orientations.

  The previous studies discussed above are all consistent with commonly held views of the role of microcracks in seismological properties. It is well known that elastic wave velocities are very sensitive to the presence of open microcracks, and that cracks play an essential role in brittle deformation of polycrystalline rocks \citep{PatersonWong}. However, previous investigations of serpentinites have focused on seismic properties during hydrostatic loading. The \emph{seismological signature} of serpentinites as they \textit{deform} under differential stress has not yet been constrained by any experimental data up to date, even though it is essential for interpreting seismic observations in active subduction zones, notably the relation between P- \textit{vs.} S-wave velocity ratios and the presence of serpentinites \citep{Reynard2013}. The signature is fundamental to monitoring and understanding the propagation of microcracks and resulting \emph{stress-induced anisotropy} in seismic wave velocities. While most polycrystalline rocks exhibit dilatancy prior to brittle failure as a result of the ``mode I'' opening of microcracks oriented parallel to the compression axis, and especially at low confining pressures \citep{PatersonWong}, the brittle deformation of antigorite at room temperature has been shown to be non-dilatant and unique in that it is accomodated entirely by shear microcracking \citep{Escartinetal1997}. However, further work is needed to clarify the dominant \emph{micromechanical mechanism} leading to brittle deformation of antigorite, notably the amount by which deformation is localised and/or distributed, and how.

  In this study, we \emph{combine mechanical and wave velocity measurements} on polycrystalline, $>95\%$ antigorite, serpentinite specimens in a triaxial apparatus, at room temperature, in order to elucidate the micromechanics accomodating deformation in antigorite. In particular, we use both elastically isotropic and anisotropic antigorite samples, allowing the investigation of stress-induced anisotropy rather than existing anisotropy, and a Westerly granite sample for comparison. Measurements of strain and ultrasonic P- and S-wave velocities were recorded during both \emph{direct} and \emph{cyclic differential loading} tests at confining pressures up to \SI{150}{\mega\pascal} on antigorite, and during one direct loading test at \SI{100}{\mega\pascal} on Westerly granite. Velocities were measured in various directions with respect to the axial stress. This combination of strain and velocity measurements allows for joint quantitative measurements of stress-strain behaviour and dilatancy, as well as the evolution of P and S-wave velocities and P-wave anisotropy during brittle deformation. ``Controlled'' or ``quasi-static'' rock failure was achieved during some cyclic loading tests on antigorite, in order to document the evolution of wave velocities \emph{during failure}. Mechanical and velocity data are reported for both direct and cyclic loading tests, as well as microstructures in recovered antigorite specimens. Discussion of the results focuses first on a comparison of P- and S-wave measurements at increasing confining pressure with literature data. The pressure dependence of P- and S-wave velocities is then intepreted in terms of crack closure, and inverted to yield quantitative estimates of crack density and aspect ratio distribution of spheroidal cracks, using the differential effective medium model of \citet{DavidRWZ2012}. The pressure dependence of dynamic and static moduli is also compared. The stress-induced anisotropy during brittle deformation is then interpreted in terms of opening of axial ``mode I'' microcracks during axial compression, and the quantity of such stress-induced microcracks is estimated for antigorite and granite, using the \citet{SayersKachanov1995} analytical expressions. The evolution of velocities \emph{during} failure, and the amount by which microcracking is localised, are discussed by looking at \emph{specific} wave velocity raypaths across, along and out of the incipient fault plane \emph{during} quasi-static antigorite rupture. A summary of the peculiar features of brittle deformation in antigorite is then given, based on joint interpretations of mechanical measurements, velocity measurements, and microstructural observations, followed by a discussion on the micromechanics of non-dilatant brittle deformation in antigorite.
  
  \section{Experimental Materials and Methods}
  \subsection{Rock Types}
  \label{subsec:spec}
\begin{figure*}
    \centering
    \includegraphics[width=0.82\textwidth]{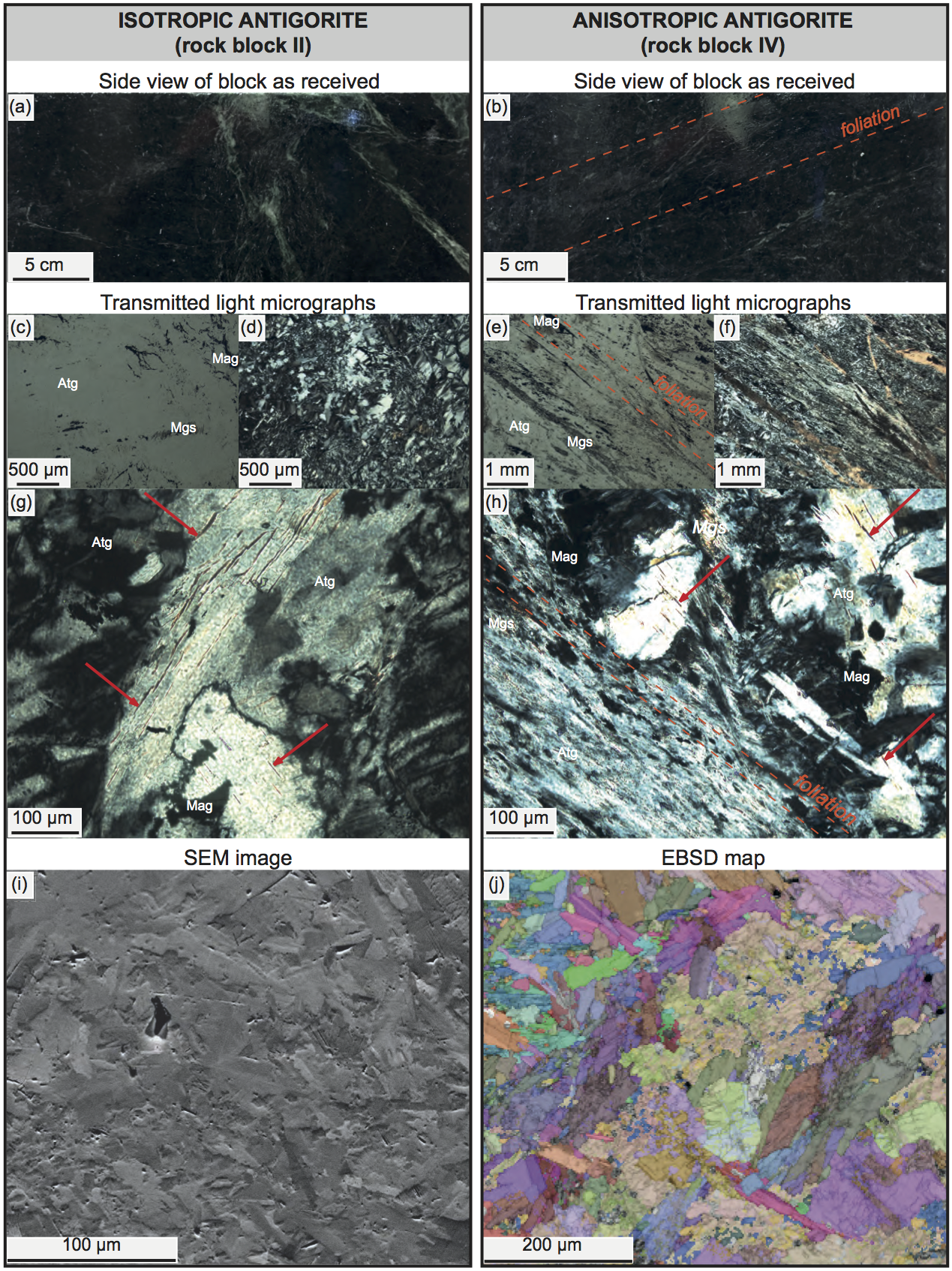}
    \caption{Representative images of the two blocks of Vermont antigorite serpentinite, as received, at various scales. a,b) Photographs of the rectangular parallelepiped blocks of serpentinite, lateral view. c,e,g) Plane-polarised transmitted light micrographs of polished thin sections; d,f,h) crossed-polarised transmitted light micrographs equivalents of c,e,g), respectively. Red arrows in g,h) indicate cracks following cleavage planes of antigorite. i) Forescatter electron image and j) Electron backscatter diffraction (EBSD) map, with orientation coloured by Euler angle and semi-transparent on top of a band contrast map. The forescatter electron image and EBSD data were acquired using Oxford Instruments AZtec software on a FEI Quanta 650 field-emission gun scanning-electron microscope (SEM), equipped with an Oxford Instruments Nordlys S EBSD camera in the Department of Earth Sciences at the University of Oxford.} 
    \label{fig:micro_asis}
  \end{figure*}

  Blocks of Vermont antigorite serpentinite (VA) (30 x 20 x 12 cm in size) were acquired from Vermont Verde Antique's Rochester quarry, Vermont, USA. This is the same material as studied by \citet{Reinenetal1994,Escartinetal1997,ChernakHirth2010}, apart from block to block variability -- see below. X-ray diffraction analysis reveals that the rock is essentially composed of pure antigorite ($>$95$\%$, dark green regions in Figure \ref{fig:micro_asis}a,b), with a minor amount of magnetite and magnesite more abundantly found within veins (white veins in Figure \ref{fig:micro_asis}a,b), both at about the 2$\%$ level. A remarkable amount of grain size and shape heterogeneity is observed in antigorite (Figure \ref{fig:micro_asis}c-j). Grain sizes are extremely heterogeneous, with abundant fine-grained regions with grain sizes typically in the range 1--10~\si{\micro\metre}, and coarser grains up to 200~\si{\micro\metre} (Figure \ref{fig:micro_asis}i,j). Grains are generally elongated with random orientations of their long axes. The pre-existing cracks that can be observed are usually found along cleavage planes following the antigorite corrugated structure (Figure \ref{fig:micro_asis}g,h). The porosity was measured using a Helium-pycnometer (Micromeritics AccuPyc II 1340), and less than $<0.1\%$ porosity (the detection limit) was detected. Rock density, calculated from dry weight measurements in cored specimens, is 2650$\pm$20~\si{\kilogram\per\cubic\metre} -- only slightly above the density of antigorite (\SI{2620}{\kilogram\per\cubic\metre} \citep{Bezacieretal2010}) and consistent with minor amounts of magnesite and magnetite.

  Two different blocks of Vermont antigorite were used in this study and characterised by P and S-wave measurements taken under ambient conditions, in multiple directions and locations, on the whole block and also on core specimens. The ``pulse-transmission'' method used for seismic velocity measurements under ambient conditions only differs from the one used for measurements under pressure (see details in section  \ref{subsubsec:expveloc}) in that large wave transducers are used (Panametrics Olympus IMS), and that arrival times are all picked manually on an oscilloscope.

  The block labelled ``block II'' has no apparent foliation, with homogeneous regions formed of pure antigorite, separated by large white veins (Figure \ref{fig:micro_asis}a). Velocity measurements are found to be very consistent if taken on the whole block and/or within core specimens, \textit{i.e.}, the velocities are scale-independent. Velocity anisotropy is less than $5\%$ (the detection limit), with P and S-wave velocities averaging 6.5 and 3.7~\si{\kilo\meter\per\second}, respectively. Accordingly, this block is considered ``isotropic''. 
  
  In contrast, the other block labelled ``block IV'' has small and regularly distributed veins, forming an apparent foliation inclined at abour \ang{20} to vertical in Figure \ref{fig:micro_asis}b. Velocity measurements are also found to be very consistent if taken on the whole block and/or within specimens cored from it, but only in one given direction. Velocity anistropy is about 15$\%$, with P and S-wave velocities ranging between 5.2--6.4 and 3.4--3.9~\si{\kilo\meter\per\second}, respectively, depending on direction. Accordingly, such rock is referred to as ``anisotropic''. Additional characterization of seismic anisotropy was done by taking P and S-wave measurements at \ang{10} intervals around a cylindrical specimen cored in the out-of-plane direction of the block as shown in Figure \ref{fig:micro_asis}b, \textit{i.e.}, in a direction having foliation parallel to the cylinder axis. The observed angular variation of P and SH-wave velocities with respect to the direction normal to the foliation (Figure \ref{fig:TI}, symbols) is adequately captured by an elastic model of \emph{transverse isotropy} (Figure \ref{fig:TI}, curves), showing maximum (minimum) velocities when wave propagation direction is parallel (perpendicular) to foliation. The direction of the foliation observed on block IV (Figure \ref{fig:micro_asis}b) matches well the orientation of the plane of transverse isotropy found from wave velocities measurements (Figure \ref{fig:TI}). The elastic tensor was inverted from the angular variation of P and SH-wave (group) velocities, by using the full analytical expressions of \citet{Thomsen1986}, to yield:
  \begin{equation}
    \label{eq:Cij}
    \mathbb{C}=\begin{bmatrix}
      110.2 & 31.2 & 27.9 & 0 & 0 & 0 \\
      31.2 & 110.2 & 27.9 & 0 & 0 & 0 \\
      27.9 & 27.9 & 72.1 & 0 & 0 & 0 \\
      0 & 0 & 0 & 31.1 & 0 & 0 \\
      0 & 0 & 0 & 0 & 31.1 & 0 \\
      0 & 0 & 0 & 0 & 0 & 39.5 \\
    \end{bmatrix} \textrm{in \si{\giga\pascal}},
  \end{equation}
  which compares qualitatively well with the set of single-crystal measurements of \citet{Bezacieretal2010} at ambient conditions. Thomsen parameters ($\epsilon$,$\gamma$,$\delta$), which are calculated from relative contributions of elastic constants \citep{Thomsen1986}, are useful dimensionless numbers to characterise the degree of anisotropy. It is found that $\epsilon$~=~0.27, $\gamma$~=~0.14 and $\delta$~=~0.31, which are indicative of a ``moderate'', nearly elliptical ($\delta \approx \epsilon$) anisotropy.

    \begin{figure*}[t]
    \centering
    \includegraphics[width=0.7\textwidth]{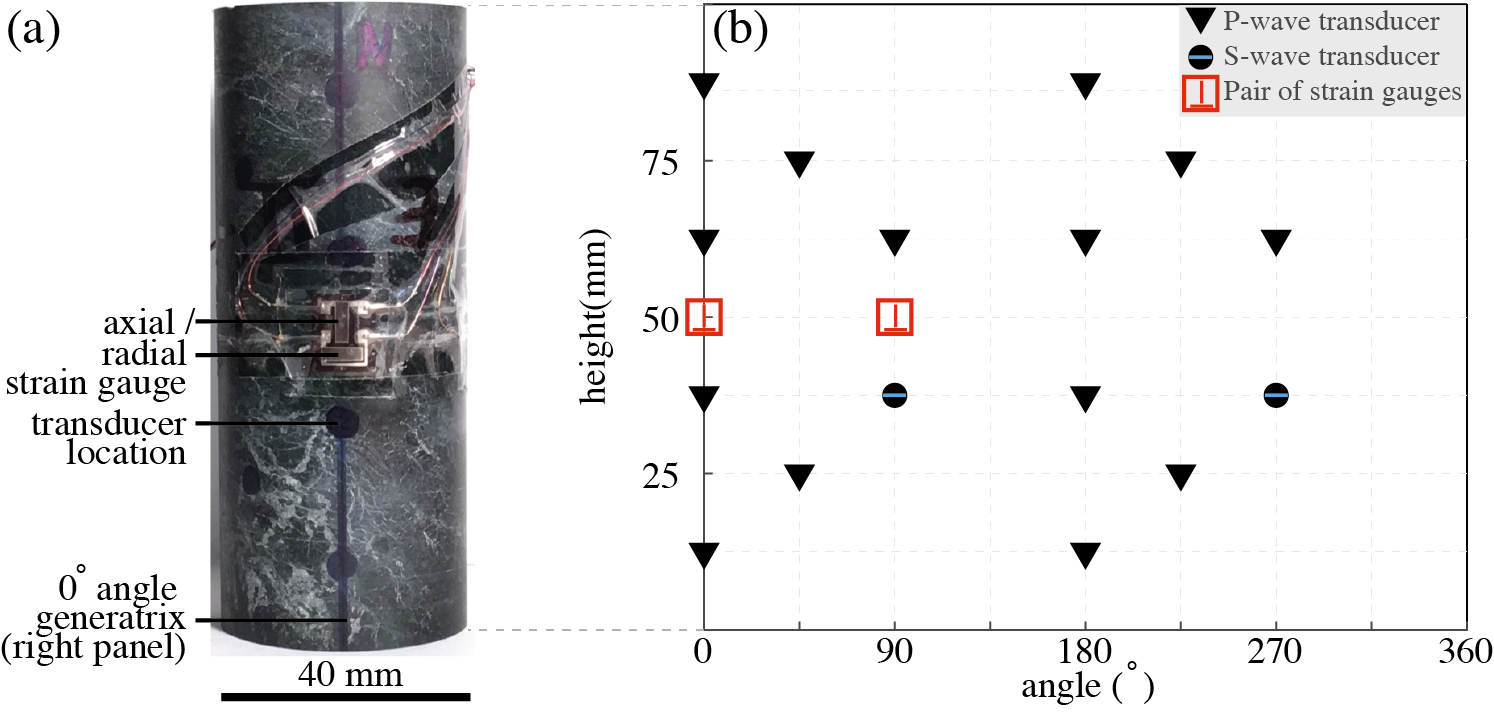}
    \caption{a) Photograph of antigorite serpentinite sample (VA -- II.3), prior to insertion in jacket, used in the UCL system. b) Map of sensor positions around the sample (2 pairs of strain gauges, 14 P- and 2 SH-wave transducers) modified from \citet{Brantutetal2014}. The blue line on the S-wave acoustic transducer symbols indicates direction of wave polarisation.} 
    \label{fig:sensor}
  \end{figure*}

  Westerly granite (WG) is used to compare our mechanical and velocity measurements with data on a polycrystalline, elastically isotropic, silicate reference material (\textit{e.g.}, \citep{Birch1960,Simmons1964,NurSimmons1969a}). This material is homogeneous, fine- to medium-grained, with grain sizes in the range 0.05--2.2~\si{\milli\metre} with an average of \SI{0.75}{\milli\metre} \citep{MooreLockner1995}. The rock is composed of 30$\%$ quartz, 35$\%$ microcline feldspar, 30$\%$ plagioclase feldspar, and 5$\%$ mica \citep{Brace1965}. Measured rock density is \SI{2.65e3}{\kilogram\per\cubic\metre}. Microcracks are observed in about 20--30$\%$ of grain boundaries \citep{SpruntBrace1974,TapponnierBrace1976} and, in similar proportions, inside grains \citep{MooreLockner1995}, with variable crack length typically less than \SI{300}{\micro\meter} \citep{MooreLockner1995}.

  \subsection{Sample Preparation}
  \label{subsec:sampleprep}
  Cylindrical specimens of 40.5$\pm$0.02 ~\si{\milli\metre} diameter and 100.0$\pm$0.02 ~\si{\milli\metre} length (see example in Figure \ref{fig:sensor}a) were cored from the two blocks of Vermont antigorite (see description above) and from a block of Westerly granite. The Vermont antigorite samples from the ``isotropic block'' (hereafter labelled as VA--II.$n$, where $n$ is sample number; see Table \ref{tab:samples}), were all cored from the vein free regions of the block. The Vermont antigorite samples from the ``anisotropic block'' (hereafter labelled as VA--IV.$n$, where $n$ is sample number; see Table \ref{tab:samples}) were cored in two different orientations with respect to the foliation: $\beta$~=~\ang{70} (foliation nearly horizontal) and $\beta$~=~\ang{0} (foliation vertical), where $\beta$ denotes the angle formed by the cylinder axis (\textit{i.e.}, the compression axis) and the foliation. The samples were then ground such that the ends were flat and parallel. Two pairs of longitudinal and radial strain gauges were glued directly onto a finely polished sample surface (Figure \ref{fig:sensor}a) using cyanoacrylate glue. The sample was then oven-dried at \SI{60}{\celsius} for at least \SI{24}{\hour} and jacketed in a viton sleeve, equipped with piezoelectric transducers (see Figure 2 of \citet{Brantutetal2014}, Figure \ref{fig:sensor}b and description below), prior to mechanical testing.

  \begin{table*}
    \caption{Summary of experimental conditions. All samples were mechanically tested at 10$^{-5}$~\si{\per\second} strain rate. $\beta$~=~\ang{0} and $\beta$~=~\ang{90} indicate foliation is vertical and horizontal, respectively, on a cylindrical sample (see section \ref{subsec:sampleprep}). $P_{\mathrm{c}}$: confining pressure. $\sigma_{\mathrm{max}}$: maximum differential stress during experiment. For details on type of test, please refer to section \ref{subsec:exptech}.}
    \label{tab:samples}
    \centering
    \begin{tabular}{|c|c|c|c|c|c||c|}
      \hline
      Sample name & $\beta$, deg & $P_{\mathrm{c}}$, \si{\mega\pascal} & $\sigma_{\mathrm{max}}$, \si{\mega\pascal} & Type of test \\
      \hline
      VA -- II.1 & none & 50 & 457\textsuperscript{$\dagger$} & cyclic loading, controlled failure \\
      VA -- II.2 & none & 100 & 639\textsuperscript{$\dagger$} & cyclic loading, controlled failure \\
      VA -- II.3 & none & 150 & 679\textsuperscript{$\dagger$} & cyclic loading, controlled failure \\
      VA -- II.4 & none & 100 & 590 & cyclic loading\textsuperscript{$\ddagger$} \\
      VA -- IV.1 & 70 & 100 & 563\textsuperscript{$\dagger$} & direct loading, failure \\
      VA -- IV.2 & 70 & 100 & 558 & cyclic loading \\
      VA -- IV.3 & 70 & 50 & 426\textsuperscript{$\dagger$} & direct loading, failure \\
      VA -- IV.4 & 70 & 50 & 373 & cyclic loading \\
      VA -- IV.5 & 70 & 150 & 674\textsuperscript{$\dagger$} & cyclic loading, controlled failure \\
      VA -- IV.6 & 70 & 20 & 316\textsuperscript{$\dagger$} & direct loading, failure \\
      VA -- IV.7 & 70 & 150 & 675\textsuperscript{$\dagger$}  & direct loading, failure \\
      VA -- IV.8 & 70 & 150 & 629 & cyclic loading \\
      VA -- IV.14 & 0 & 50 & 378\textsuperscript{$\dagger$} & cyclic loading, controlled failure \\
      VA -- IV.15 & 0 & 100 & 507\textsuperscript{$\dagger$} & cyclic loading, controlled failure \\
      WG2 & none & 100 & 765\textsuperscript{$\dagger$} & direct loading, failure \\
      \hline
    \end{tabular}
    \\
    \flushleft
    \textsuperscript{$\dagger$}{\footnotesize{Denotes also rock strength or ``peak stress'' (associated with rock failure)}}\\
    \textsuperscript{$\ddagger$}{\footnotesize{Multiple cycles at same stress}}
  \end{table*}

  \subsection{Experimental Technique and Data Analysis}
  \label{subsec:exptech}
  \subsubsection{Triaxial Deformation Experiments}
  \label{subsubsec:exptriax}
  Deformation experiments were conducted in an oil medium triaxial apparatus at the Rock and Ice Physics Laboratory at University College London (see description in \citet{Ecclesetal2005}). The apparatus allows independent application of servo-controlled confining pressure (by a hydraulic pump) and autocompensated axial load (by a hydraulic actuator and piston). Improved corrections for \emph{piston friction} were made to accurately extract the true load on the rock sample from the load measured \emph{externally} by a load cell, at various conditions of confining pressure and axial load. Such corrections, details of which are given in Appendix A, are particularly important when piston direction is reversed multiple times in cylic loading measurements (see below), and were greatly facilitated by the use of strain gauges on the sample which allows monitoring of any change of stress on rock samples.

  Table \ref{tab:samples} summarises samples and experimental conditions. Samples were initially loaded hydrostatically to the desired confining pressure, $P_{\textrm{c}}$, using \SI{10}{\mega\pascal} pressure steps and $\sim$\SI{15}{\minute} dwell times, and then axially deformed at a strain rate of 10$^{-5}$~\si{\per\second}. Two types of deformation experiments were performed: \emph{direct loading} tests until rock failure, and \emph{cyclic loading} tests with multiple load-unload cycles. Cyclic loading tests were conducted at increasing maximum differential stress between successive cycles, except one ``fatigue test'' with multiple cycles at the same differential stress at $P_{\mathrm{c}}$~=~\SI{100}{\mega\pascal} (see Table \ref{tab:samples}). After deformation, hydrostatic pressure was decreased using \SI{10}{\mega\pascal} pressure steps and $\sim$\SI{5}{\minute} dwell times. Recovered specimens were embedded in epoxy and sectioned for preparation of polished thin-sections for optical and scanning-electron microscopy.

  \paragraph{Quasi-static, ``controlled'' failure tests}
  Direct loading tests at constant strain rate result in a violent rock failure, associated with a large axial stress drop and occurring in a fraction of a second. In order to keep slow strain rates and obtain velocity measurements \emph{during} rock failure, cyclic loading tests were also used to achieve ``controlled'' or ``quasi-static'' failure, by applying small amplitude load-unload cycles while approaching failure.  Although the system used at University College London allows for passive acoustic emission monitoring in conjunction with active ultrasonic velocity surveys (\textit{e.g.}, \citep{Brantut2018}), no acoustic emissions could be detected in the antigorite deformation experiments. For this reason, the method of controlled failure using feedback from acoustic emission originally proposed by \citet{Lockneretal1991} could not be implemented. The method employed here was similar to that of \citet{WawersikBrace1971}: guided by direct observation of \emph{stress-strain data} and, more precisely, by the instability of strain gauge data when strain becomes localised (see below), axial stress was adjusted \emph{manually} by the operator to maintain a stable mechanical behaviour in the post-failure region.

  \subsubsection{Rock Strain Measurements}
  \label{subsubsec:expstrain}
  The axial shortening is measured with a pair of external linear variable differential transformers (LVDTs), and corrected from machine stiffness to measure rock specimen axial strain. In addition to LVDT measurements, rock strain gauge measurements were implemented in the triaxial apparatus during this study. Specimens were equipped with two pairs of axial and circumferential electric resistance strain gauges (Tokyo Sokki TML--FCB), located at \ang{90} around the specimen axis (Figure \ref{fig:sensor}). Each \SI{350}{\ohm} resistance was mounted on a precision Wheatstone quarter-bridge. Strain gauges provide \emph{local} measurements of both axial ($\epsilon{_{\mathrm{ax}}}$) and circumferential strain ($\epsilon{_{\mathrm{circ}}}$), with high precision ($\sim$10$^{-6}$). The volumetric strain $\epsilon{_{\mathrm{v}}}$ can be directly estimated as
  \begin{equation}
    \label{eq:epsilonvol}
    \epsilon{_{\mathrm{v}}}=\frac{1}{2}(\epsilon_{\mathrm{ax}}^{(1)}+\epsilon_{\mathrm{ax}}^{(2)})+\epsilon_{\mathrm{circ}}^{(1)}+\epsilon_{\mathrm{circ}}^{(2)},
  \end{equation}
  where the superscript denotes a given pair of axial and circumferential strain gauges. Local strain gauges measurements are more likely to capture the process of strain localisation in rocks, while approaching failure, than external LVDTs measurements -- a property advantageously used during ``controlled failure'' tests (see above). However, quantitative use of such data is inherently limited during strain localisation, which leads to unstable, divergent (and sometimes non-monotonic) strain gauge signals, and eventually to strain gauge failure. An example of raw data, showing axial and circumferential strain gauge measurements on the two pairs of strain gauges, and axial strain measurements from LVDTs, is given in Figure \ref{fig:mecaraw_direct} for direct loading test on antigorite at \SI{150}{\mega\pascal} confining pressure. Axial and radial strain measurements are consistent between the two pairs of strain gauges. In addition, axial strain measurements from both the LVDTs and local strain gauges show a very good agreement, except just prior to failure where strain becomes localised as previously noted. For a few experiments, based on the comparison above, some strain gauge data were discarded (\textit{e.g.}, caused by poor gluing or electrical noise). Note that the convention used for strain measurements is that positive strains are compressive.

  \subsubsection{Wave Velocity Measurements}
  \label{subsubsec:expveloc}
  In conjunction with rock strain measurements, active ultrasonic velocity surveys were performed (every minute) using 14 P- and 2 SH-wave ultrasonic piezoelectric transducers mounted around the rock specimen (Figure \ref{fig:sensor}b). A high voltage pulse ($\sim$ \SI{250}{\volt}) is successively sent to each transducer, at central frequency of \SI{1}{\mega\hertz}, producing a mechanical wave at known origin time. The received waveforms are recorded by the remaining sensors. The precise P- or S- wave arrival times are extracted from the waveform data using a recently improved cross-correlation technique described in \citet{Brantutetal2014}, with reference to a ``master survey'' where arrival times are picked manually. The wave velocity is directly obtained by dividing the distance between a given pair of transducers (corrected from rock strain data) by the the arrival time (corrected for time of flight in the metal-support piece of the wave transducer). Considering all uncertainties arising from the manual arrival time picking process, calibration of traveltime in transducer metal ends, and travel distance, P- and S-wave velocities are accurate within 3 and 5$\%$, respectively; but after cross-correlation the relative precision between successive velocity measurements is as high as 0.2$\%$.

  The selected raypaths are the ones intersecting the axis of the sample, a condition which is equivalent to a \ang{180} spacing between a pair of transducers in Figure \ref{fig:sensor}b. The geometrical arrangement of sensors allows for measurements of P-wave velocities at four different angles with respect to the axial stress: \ang{90}, \ang{58}, \ang{39} and \ang{28} on each of the 7, 6, 6, and 4 pairs of P-wave transducers, respectively. Except when stated otherwise, the P-wave velocity data reported in this paper always correspond to the \emph{average} of P-wave velocities over all available raypaths, at a given angle to the compression axis. SH-wave velocities are measured at \ang{90} to the compression axis on one pair of S-wave transducers. The direction of polarisation of the SH-waves is shown in Figure \ref{fig:sensor}b. 

  \begin{figure*}[b]
    \centering
    \includegraphics[width=\textwidth]{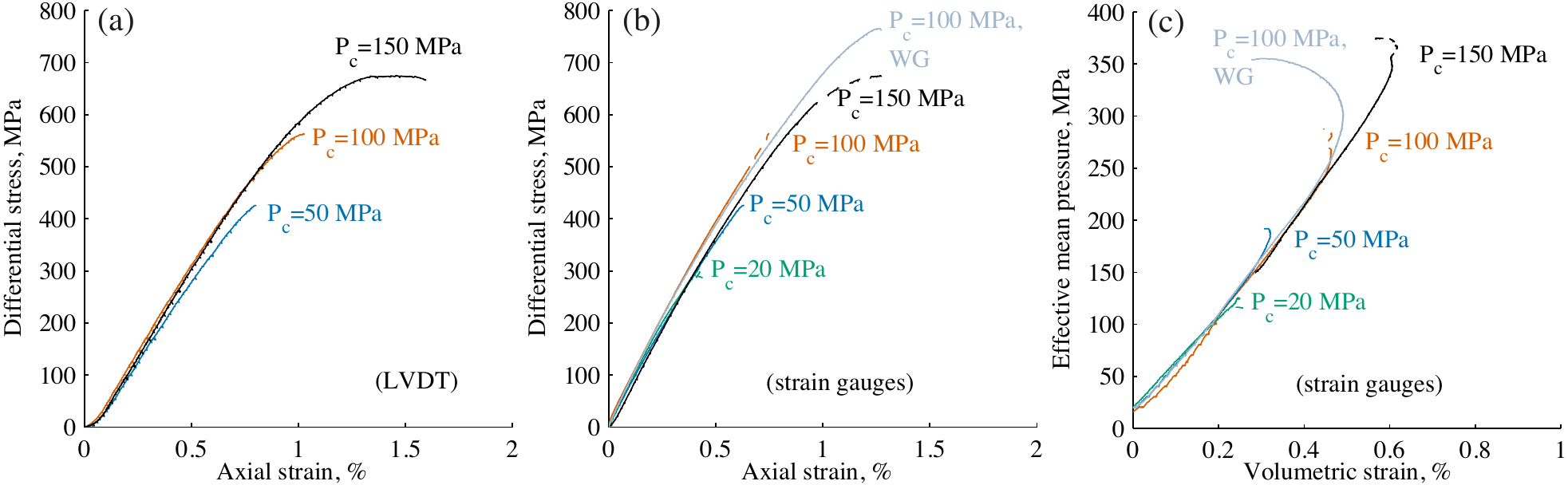}
    \caption{Stress-strain data for all direct loading tests until rock failure (see Table \ref{tab:samples}), on antigorite at various confining pressures $P_{\textrm{c}}$, and on Westerly granite (WG) at $P_{\textrm{c}}$~=~\SI{100}{\mega\pascal}. a) Axial strain (measured by LVDTs) \textit{vs.} differential stress curves. Data for samples VA-IV-06 and WG2 are not available due to technical issues. b) Axial strain (measured by strain gauges) \textit{vs.} differential stress curves. c) Volumetric strain (calculated from strain gauge measurements) \textit{vs.} effective mean stress curves. Dashed lines indicate final portion of the stress-strain curve where strain gauge data diverge from each other and from LVDT measurements (\textit{e.g.}, Figure \ref{fig:mecaraw_direct}), interpreted as strain localisation. For details on strain measurements, see section \ref{subsubsec:expstrain}.}
    \label{fig:meca_direct}
  \end{figure*}
  
  \section{Results}
  \label{sec:resultsdirect}
  \subsection{Mechanical Data}
  \label{subsec:res_mech}
  \subsubsection{Direct Loading Experiments}
  \label{subsubsec:res_mech_direct}
  The axial stress \textit{vs.} differential stress curves for \emph{direct loading tests} (Figures \ref{fig:meca_direct}a,b) reveal that the behaviour of polycrystalline antigorite samples is linearly elastic up over a large stress range, typically $\gtrsim$80$\%$ of rock strength. Deviation from linear elastic behaviour, \textit{i.e.}, yield point, occurs close to rock failure, which is abrupt and associated with a large stress drop. Such ``elastic-brittle'' behaviour is observed at all confining pressures $\leq$\SI{150}{\mega\pascal}, although a minor amount of strain weakening prior to rock failure is observed at $P_{\mathrm{c}}$~=~\SI{150}{\mega\pascal}. The behaviour of the Westerly granite sample is also brittle at $P_{\mathrm{c}}$~=~\SI{100}{\mega\pascal}, however, the yield point in granite occurs at about 50$\%$ of rock strength, which is a much lower percentage than observed for antigorite.

  At $P_{\mathrm{c}}$~=~\SI{100}{\mega\pascal}, antigorite strength is only about 25$\%$ less than that of Westerly granite (Figure \ref{fig:meca_direct}b). The differential stress at brittle failure for antigorite ($\sigma_{\mathrm{max}}$, Table \ref{tab:samples}), follows a Coulomb failure criterion. $\sigma_{\mathrm{max}}$ increases linearly with confining pressure (Figures \ref{fig:meca_direct}a,b) as $\sigma_{\mathrm{max}}=305+3.59P_{\mathrm{c}}$ if values are expressed in \si{\mega\pascal} (fit over all rock failure data of Table \ref{tab:samples}, except samples VA--IV.14\&15). If written in terms of the normal ($\sigma_{\textrm{n}}$) and shear stress ($\tau$) acting on the failure plane, this criterion is expressed as $\tau=81+0.68\sigma_{\textrm{n}}$, where the second coefficient on the r.h.s. of this equation is the coefficient of static friction. Although experiments in this study are limited to \SI{150}{\mega\pascal} confining stress, the failure criterion and high values of antigorite strength are in excellent agreement with previous experimental results of \citet{RaleighPaterson1968} and \citet{Escartinetal1997}.

  The volumetric strain \textit{vs.} effective mean stress curves for direct loading tests are shown in Figure \ref{fig:meca_direct}c. The initial part of these curves -- until the value of effective mean stress reaches the confining pressure, for a given experiment -- corresponds to hydrostatic compression, followed by axial deformation. For all samples, volumetric strain increases approximately linearly with effective mean stress, which corresponds to elastic compression. Potential deviation from this elastic ``reference'' line towards negative volumetric strain defines the onset of dilatancy \citep{PatersonWong}, \textit{i.e.}, inelastic volume increase. The dilatancy observed prior to failure on antigorite samples at all confining pressures is negligible ($<$0.05$\%$ volumetric strain), particularly if compared to Westerly granite ($\approx$0.3$\%$ volumetric strain at $P_{\textrm{c}}$~=~\SI{100}{\mega\pascal}, Figure \ref{fig:meca_direct}c). The onset of (any) dilatancy in antigorite is observed very close to failure, at all confining pressures. However, detection of dilatancy at such late stages prior to failure could be biased by the effect of strain localisation on local strain gauge measurements as described above \citep[see also][]{Escartinetal1997}. In contrast, Westerly granite starts dilating at about 50-60$\%$ of failure stress (at $P_{\textrm{c}}$~=~\SI{100}{\mega\pascal}, Figure \ref{fig:meca_direct}c), as previously reported in many previous experimental studies (\textit{e.g.}, \citet{Braceetal1966}). The comparative test on Westerly granite also demonstrates that any noticeable dilatancy in antigorite specimens would have been measured by the strain gauges.  The ``non-dilatant'' particular character of antigorite brittle deformation observed in this study is consistent with results first reported by \citet{Escartinetal1997} at similar confining pressures.

  \begin{figure*}[b]
    \centering
    \includegraphics[width=0.82\textwidth]{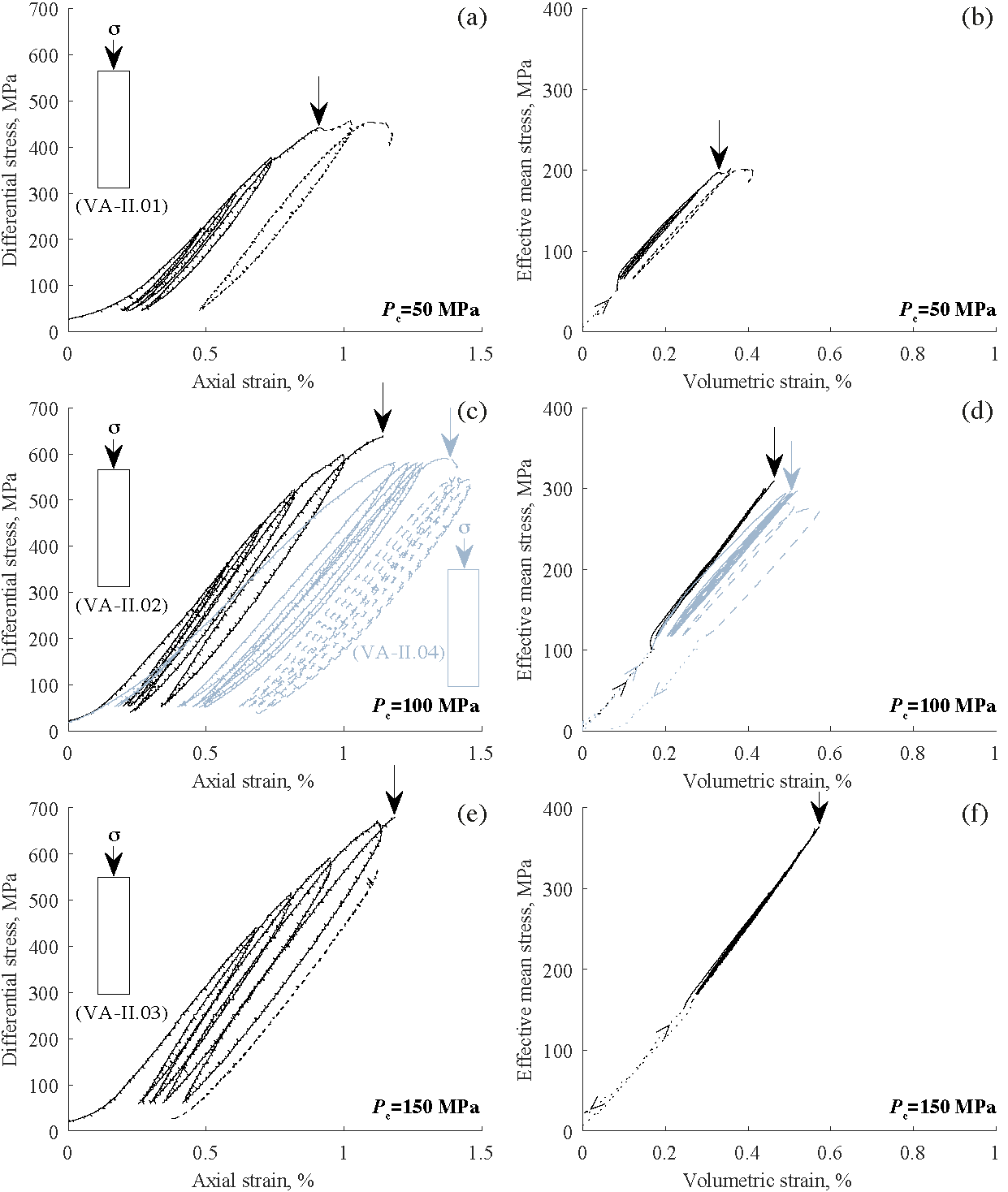}
    \caption{Stress-strain data for all cyclic loading tests on the \emph{isotropic antigorite} samples (see Table \ref{tab:samples}) at various confining pressures: (a,b) $P_{\textrm{c}}$~=~\SI{50}{\mega\pascal}; (c,d) $P_{\textrm{c}}$~=~\SI{100}{\mega\pascal}; (e,f) $P_{\textrm{c}}$~=~\SI{150}{\mega\pascal}. Dotted lines: hydrostatic compression or decompression (the small arrow indicates sense of pressure variation); full lines: axial load-unload cycles. Arrows on curves indicate \emph{quasi-static rock failure} (see section \ref{subsubsec:exptriax} for details), and dashed lines the portion of stress-strain curve after strain has localised. (a,c,e) Axial strain (from LVDT) \textit{vs.} differential stress data; (b,d,f) Volumetric strain (from strain gauges) \textit{vs.} effective mean stress data.} 
    \label{fig:meca_cyclicII}
  \end{figure*}

  \subsubsection{Cyclic Loading Experiments}
  \label{subsubsec:res_mech_cyclic}
  The mechanical data for \emph{cyclic loading tests} on isotropic antigorite samples, at confining pressures of 50, 100 and 150~\si{\mega\pascal}, are shown in Figure \ref{fig:meca_cyclicII} (for sample details, see Table \ref{tab:samples}). At $P_{\textrm{c}}$~=~50~\si{\mega\pascal}, the behaviour of the antigorite sample is purely elastic during the first and second cycles, to a maximum differential stress of 200 and 300~\si{\mega\pascal}, respectively (Figure \ref{fig:meca_cyclicII}a). Note that, at small stress and particularly during the first cycle, the initial ``concave-upward'' slope in the axial strain-stress curve is mostly an experimental artefact related to the initial compression of elements in the loading column and frictional effects, which is well identified by comparing axial strain data measured by LVDTs with those from local strain gauge measurements (not shown here). Accordingly, such behaviour is not observed in the second cycle at small stress, and the rock approximately follows the same stress-strain curve during loading and unloading. A small deviation from linear elastic behaviour is observed at about \SI{320}{\mega\pascal} differential stress during the third cycle to a maximum differential stress of \SI{380}{\mega\pascal}, giving rise to hysteresis in the axial strain \textit{vs.} differential stress curve, and a small permanent inelastic strain (of about 0.1$\%$) when the axial load is removed. The amount of ``permanent'' inelastic strain increases during the fourth cycle to a maximum differential stress of \SI{460}{\mega\pascal}, where a small stress drop is identified as the onset of rock failure. The observation that the strength of the rock has been reached during this cycle is confirmed by subsequent reloading during the final cycle (Figure \ref{fig:meca_cyclicII}a). The corresponding volumetric strain \textit{vs.} effective mean stress curve at $P_{\textrm{c}}$~=~50~\si{\mega\pascal} is shown in Figure \ref{fig:meca_cyclicII}b. The volumetric strain \textit{vs.} effective mean stress data follow approximatively the same slope during hydrostatic compression (dotted line) and axial deformation prior to failure (full line), even during cycles where permanent inelastic axial strain is created, suggesting that inelastic behaviour is purely non-dilatant. After rock failure (last cycle, dashed line), even if quantitative use of post-failure strain gauge data is limited as previously mentioned (see section \ref{subsubsec:expstrain}), the volumetric strain \textit{vs.} effective mean stress behaviour still compares very well with that observed during hydrostatic compression, and axial deformation cycles. Strain gauges broke soon after rock failure, which is common even in quasi-static failure tests.

  The results of cyclic loading tests on the isotropic antigorite samples at confining pressures of 100~\si{\mega\pascal} (Figures \ref{fig:meca_cyclicII}c,d) and 150~\si{\mega\pascal} (Figures \ref{fig:meca_cyclicII}e,f) show very strong similarities with the the results described above at 50~\si{\mega\pascal} confining pressure. Hysteresis is again observed only in the axial strain \textit{vs.} differential stress curves (Figures \ref{fig:meca_cyclicII}c,e) and not in the volumetric strain \textit{vs.} effective mean stress curves (Figures \ref{fig:meca_cyclicII}d,f). At a given confining pressure, the ``permanent'' inelastic axial strain (accumulated after each cycle) increases with increasing maximum differential stress but, at a given maximum differential stress, does not seem to depend on confining pressure (Figures \ref{fig:meca_cyclicII}a,c,e). During multiple cycles at the same maximum differential stress (at $P_{\mathrm{c}}$~=~\SI{100}{\mega\pascal} and about 95$\%$ or rock strength, Figure \ref{fig:meca_cyclicII}c), the ``permanent'' inelastic axial strain that is created increases slightly during successive cycles but seems to become independent from the number of cycles after three to four cycles.

      \begin{figure*}[t]
    \centering
    \includegraphics[width=0.82\textwidth]{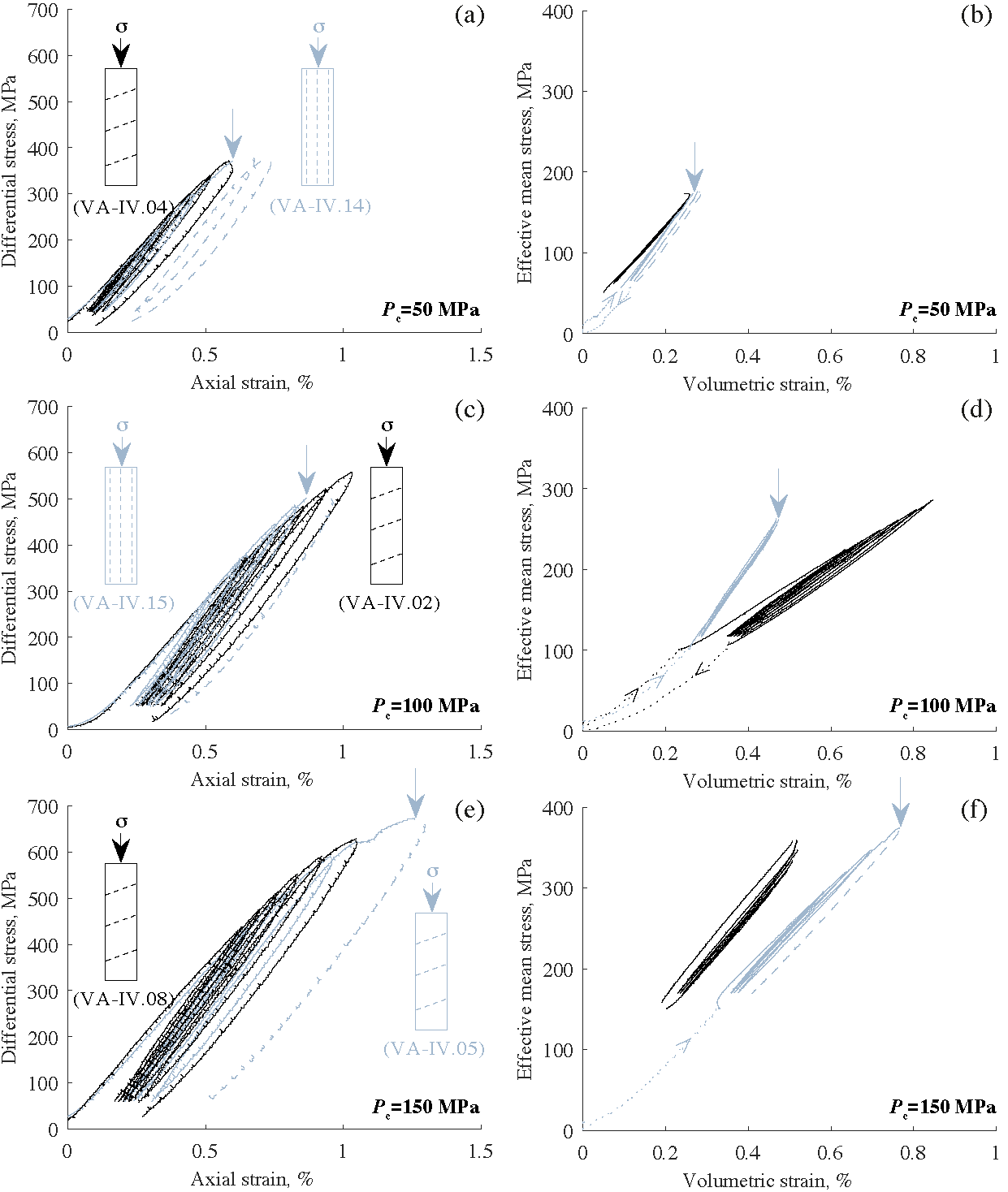}
    \caption{Stress-strain data for all cyclic loading tests on the \emph{anisotropic antigorite} samples (see Table \ref{tab:samples}) at various confining pressures: (a,b) $P_{\textrm{c}}$~=~\SI{50}{\mega\pascal}; (c,d) $P_{\textrm{c}}$~=~\SI{100}{\mega\pascal}; (e,f) $P_{\textrm{c}}$~=~\SI{150}{\mega\pascal}. Dotted lines: hydrostatic compression or decompression (the small arrow indicates sense of pressure variation); full lines: axial load-unload cycles. Arrows on curves indicate \emph{quasi-static rock failure} (see section \ref{subsubsec:exptriax} for details), and dashed lines the portion of stress-strain curve after strain has localised. (a,c,e) Axial strain (from LVDT) \textit{vs.} differential stress data; (b,d,f) Volumetric strain (from strain gauges) \textit{vs.} effective mean stress data.} 
    \label{fig:meca_cyclicIV}
  \end{figure*}

  The mechanical data for cyclic loading tests for the anisotropic samples, at confining pressures of 50, 100 and 150~\si{\mega\pascal}, are shown in Figure \ref{fig:meca_cyclicIV} (for sample details, see Table \ref{tab:samples}). In addition to samples where the foliation forms an angle of \ang{70} with direction of loading, two samples where foliation is parallel to the direction of loading have also been tested at confining pressure of 50~\si{\mega\pascal} (Figure \ref{fig:meca_cyclicIV}a,b) and 100~\si{\mega\pascal} (Figure \ref{fig:meca_cyclicIV}a,b). The anisotropic antigorite samples exhibit very strong similarities with the results described above for the isotropic samples (Figures \ref{fig:meca_cyclicII}). The inelastic behaviour is again purely non-dilatant at all confining pressures, and for both orientations of sample foliation with respect to the axial stress.

  In addition to very comparable stress-strain behaviour, values of \emph{strength} for anisotropic antigorite samples is only about 10-20$\%$ less than that of the isotropic antigorite samples (Table \ref{tab:samples}, and Figures \ref{fig:meca_direct}, \ref{fig:meca_cyclicII} and \ref{fig:meca_cyclicIV}). The strength of samples where foliation is vertical is about 10$\%$ less than that of samples where foliation is subhorizontal, at both 50 and 100~\si{\mega\pascal} confining pressure (Table \ref{tab:samples}), which is entirely consistent with the results of \citet{Escartinetal1997} (at $P_{\mathrm{c}}$~=~\SI{200}{\mega\pascal}, Figure 5 of that publication). Such results, along with the overall very comparable stress-strain behaviour for all antigorite samples, demonstrate that the presence of the foliation does not play a significant role in the mechanical behaviour.

  \subsection{Velocity Data}
  \label{subsec:res_veloc}
  \subsubsection{Direct Loading Experiments}
  \label{subsubsec:res_veloc_direct}
\begin{figure*}[b]
    \centering
    \includegraphics[width=0.83\textwidth]{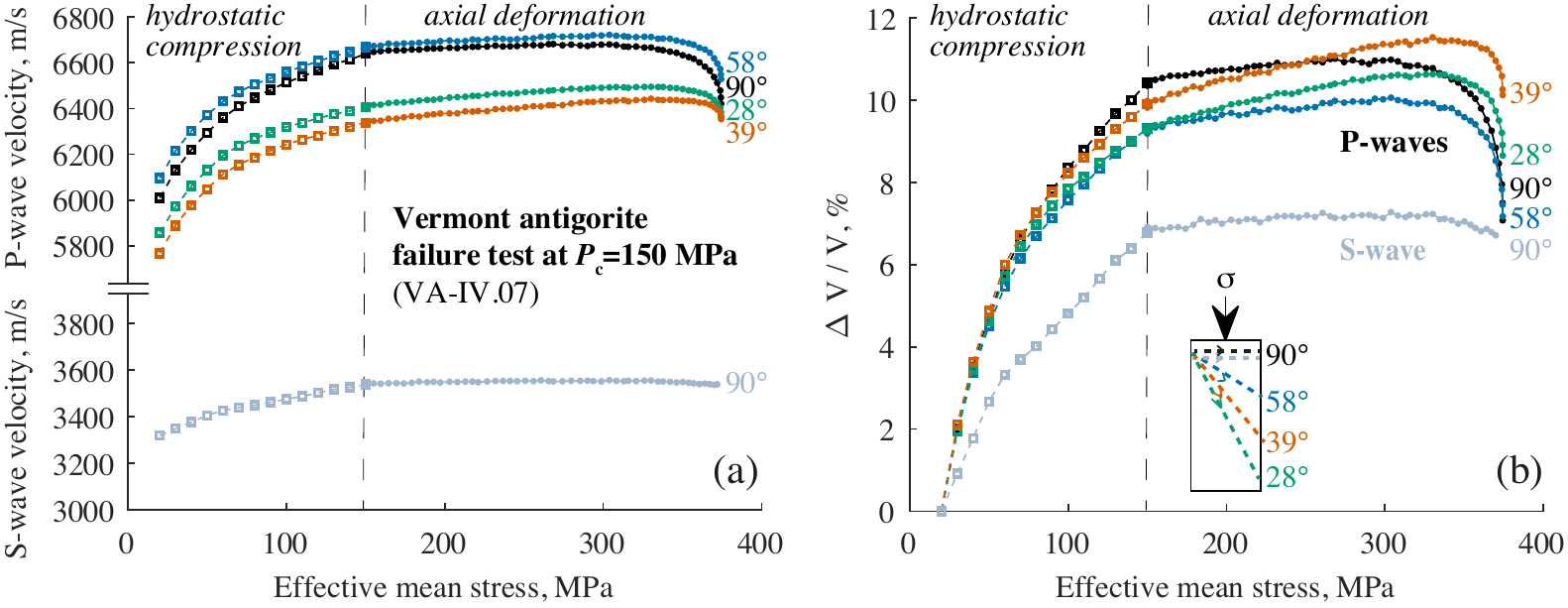}
    \caption{P and S-wave velocity data for direct loading test on antigorite until rock failure at $P_{\mathrm{c}}$~=~\SI{150}{\mega\pascal} (specimen VA--IV.07). Open squares: hydrostatic loading; full circles: axial loading. The angle is taken between direction of axial compression and ray path. a) P- and S-wave velocities, absolute values; b) P- and S-wave velocities, change relative to value at the beginning of hydrostatic compression, at $P_{\mathrm{c}}$~=~\SI{20}{\mega\pascal}. For details on velocity measurements, see section \ref{subsubsec:expveloc}.} 
    \label{fig:velocraw_direct}
  \end{figure*}

  An example of P- and S-wave velocity data is given in Figure \ref{fig:velocraw_direct} for the direct axial loading test on antigorite at $P_{\textrm{c}}$~=~\SI{150}{\mega\pascal}. Under hydrostatic compression, both P and S-wave velocities increase noticeably with increasing confining pressure in a typical ``concave-downward'' fashion by about 10$\%$ and 7$\%$, respectively. Velocities continue increasing at $P_{\textrm{c}}$~=~\SI{150}{\mega\pascal}, and do not seem to reach a plateau. The variation of both P- and S-wave velocities is much less during axial deformation (up to failure) than during hydrostatic compression. To about 80-90$\%$ of failure stress, the velocity of P-waves at \ang{90} and \ang{58} to the axial stress, and of S-waves at \ang{90} to the axial stress, is constant, while the velocity of P-waves at \ang{39} and \ang{29} to the axial stress only increases by only 1$\%$. A small decrease in both P- and S-wave velocities (by at most 3$\%$ and 0.5$\%$, respectively) occurs very close to rock failure. In addition, the small decrease in P-wave velocities during axial deformation and prior to failure is overall the same (2 to 3$\%$) in all directions with respect to the axial stress, which suggests that the brittle deformation and failure of antigorite is associated with \emph{an absence of stress-induced anisotropy}.

\begin{figure*}[b]
    \centering
    \includegraphics[width=0.82\textwidth]{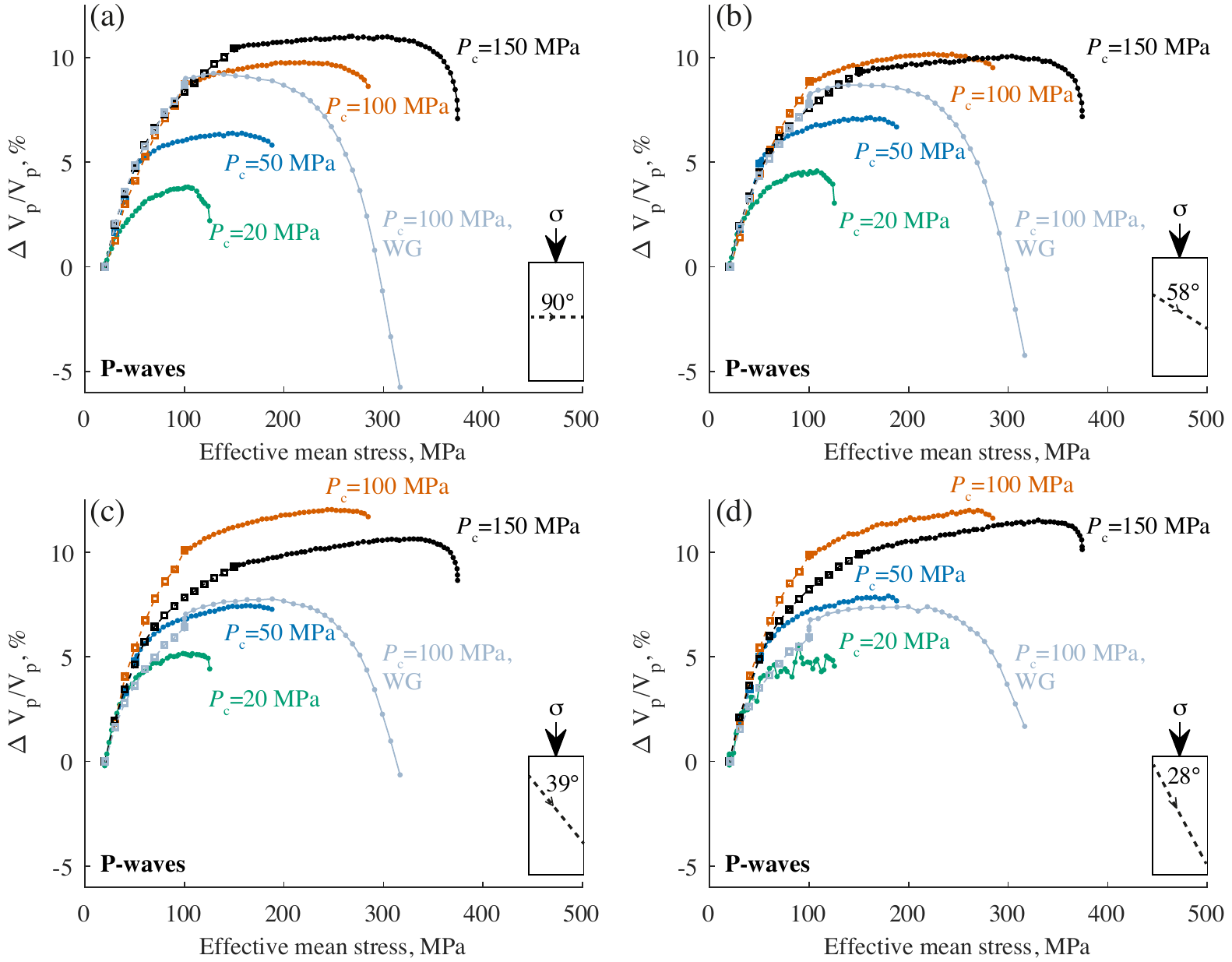}
    \caption{P-wave velocity data (relative change) for all direct loading tests until rock failure, on antigorite at various confining pressures $P_{\textrm{c}}$, and Westerly granite (WG) at $P_{\textrm{c}}$~=~\SI{100}{\mega\pascal} (see Table \ref{tab:samples}), at four angles with respect to the compression axis: a) \ang{90}; b) \ang{58}; c) \ang{39}; d) \ang{28}. Open squares: hydrostatic loading; full circles: axial loading.} 
    \label{fig:velocP_direct}
  \end{figure*}

  The relative change of P- and S-wave velocities with effective mean stress for all direct loading tests on antigorite, in the 20--150~\si{\mega\pascal} confining pressure range, is shown in Figures \ref{fig:velocP_direct} and \ref{fig:velocS_direct}, respectively. At confining pressures $\leq$\SI{100}{\mega\pascal}, the evolution of both P- and S-wave velocities during axial deformation is very similar to that described above at $P_{\mathrm{c}}$~=~\SI{150}{\mega\pascal}. The variation of P and S-wave velocities is less than 2$\%$, and for P-waves, is very similar in all directions with respect to the axial stress, again showing an absence of stress-induced anisotropy. Any decrease of wave velocities also occurs very close to failure.

  The variation of P- and S-wave velocities, for the comparative direct loading test on Westerly granite at $P_{\mathrm{c}}$~=~\SI{100}{\mega\pascal}, is also shown in Figures \ref{fig:velocP_direct} and \ref{fig:velocS_direct}, respectively. The increase of both P and S-wave velocities with increasing confining pressure, and with effective mean stress in the early part of axial deformation, is broadly comparable in both granite and antigorite. However, in granite, both P- and S-wave velocities start decreasing during axial loading at about 40-50$\%$ of failure stress, and continue to decrease until rock failure. The marked decrease in P-wave velocities evolves from 15$\%$ to 5$\%$ if taken from a direction perpendicular (Figure \ref{fig:velocP_direct}a) to nearly parallel (Figure \ref{fig:velocP_direct}d) to the compression axis, giving rise to pronounced stress-induced anisotropy. Such results on Westerly granite, which are consistent which those reported in many previous experimental studies (\textit{e.g.}, \citet{Lockneretal1977,Sogaetal1978}), highlight the remarkable absence of stress-induced anisotropy and velocity variation that is observed during direct brittle deformation in antigorite.

  \subsubsection{Cyclic Loading Experiments}
  \label{subsubsec:res_veloc_cyclic}
  The P- and S-wave velocity data for cyclic loading tests are best analysed by displaying their evolution as functions of time, and relative to their value at the beginning of the load-unload cycles (\textit{i.e}, at the given confining pressure of each test), along with differential stress as a function of time for comparison. The relative change in P- and S-wave velocities for the isotropic antigorite samples, at confining pressures of 50, 100 and 150~\si{\mega\pascal}, is shown in Figure \ref{fig:veloc_cyclicII} (for sample details, see Table \ref{tab:samples}). At $P_{\textrm{c}}$~=~50~\si{\mega\pascal} (Figure \ref{fig:veloc_cyclicII}a,b), during each cycle, P- and S-wave velocities increase during axial loading, and decrease during unloading. Until rock failure (during the fourth cycle), the change in velocity during each cycle is very small (<1.5$\%$ and <0.5$\%$ for P- and S-wave velocities, respectively), and depends weakly upon \emph{stress amplitude}. Similarly to direct loading tests (section \ref{subsubsec:res_veloc_direct}), the overall evolution in P-wave velocities is the same in all directions with respect to the axial stress prior to failure, which again indicates an absence of stress-induced anisotropy. A small decrease in both P- and S-wave velocities ($\approx$2$\%$ and 1$\%$, respectively) is observed during quasi-static rock failure, and is again associated with the absence of stress-induced anisotropy (Figure \ref{fig:veloc_cyclicII}a).

  The evolution of P- and S-wave velocities on the isotropic antigorite samples during cyclic loading tests at $P_{\textrm{c}}$~=~100~\si{\mega\pascal} (Figures \ref{fig:veloc_cyclicII}c--f) and $P_{\textrm{c}}$~=~150~\si{\mega\pascal} (Figures \ref{fig:veloc_cyclicII}g,h) yields results similar to that described above at $P_{\textrm{c}}$~=~50~\si{\mega\pascal}. In some cases (\textit{e.g.}, P-wave velocity at \ang{90} to axial stress, Figure \ref{fig:veloc_cyclicII}c), the change in velocity during load-unload cycles is so small that it approaches the limit in relative precision between successive velocity measurements (0.2$\%$, see section \ref{subsubsec:expveloc}). Also, note that application of four successive load-unload cycles at 95$\%$ of failure stress (prior to failure) during the ``fatigue'' test at $P_{\textrm{c}}$~=~100~\si{\mega\pascal} (Figures \ref{fig:veloc_cyclicII}e,f) results only in a small decrease of velocities over successive cycles, at the same stress (<0.5$\%$ in all directions with respect to direction of loading).

  The variation of P- and S-wave velocities during cyclic loading tests on the isotropic antigorite samples, at confining pressures of 50, 100 and 150~\si{\mega\pascal}, is shown in Figure \ref{fig:veloc_cyclicIV} (for sample details, see Table \ref{tab:samples}). These results are consistent with those described above for isotropic samples, in that prior to failure the variation of P- and S-wave velocities during cycles is very small (<2$\%$ and <1$\%$ for P- and S-wave velocities, respectively) and associated with no stress-induced anisotropy, at all confining pressures. This is observed for both samples in which foliation is oriented at either \ang{70} to the compression axis or parallel to it. Some cyclic loading experiments were ended without reaching rock failure (Figures \ref{fig:veloc_cyclicIV}a--b, \ref{fig:veloc_cyclicIV}e--f, and \ref{fig:veloc_cyclicIV}i--j at $P_{\textrm{c}}$~=~50, 100 and 150~\si{\mega\pascal}, respectively), at a maximum differential stress typically about 90-95$\%$ of failure stress (if rock strength is taken from direct loading tests at given confining pressure; see Table \ref{tab:samples}). For such tests, and all confining pressures (Figures \ref{fig:veloc_cyclicIV}a,e,i), wave velocities return to within 1$\%$ of their initial value at the end of the last cycle after unloading.

  \subsection{Elastic Moduli in Antigorite from Static and Dynamic Measurements}
  \label{subsec:res_staticdynamic}

  It is useful to compare values of elastic moduli that can be directly extracted from both mechanical and velocity measurements (Figure \ref{fig:staticdynamic}a). At a given confining pressure, the linear portions of the axial strain \textit{vs.} differential stress, and circumferential \textit{vs.} axial strain, yield Young's modulus, $E=\epsilon{_{\mathrm{ax}}}/ \sigma$, and Poisson's ratio, $\nu=-\epsilon{_{\mathrm{circ}}}/\epsilon{_{\mathrm{ax}}}$, respectively, where $\sigma$ denotes differential stress. In isotropic samples, the bulk and shear moduli ($K$,$G$) have been calculated using the classical elasticity relations $K=E/[3(1-2\nu)]$ and $G=E/[2(1+\nu)]$, respectively, at three confining pressures (Figure \ref{fig:staticdynamic}a). Dynamic bulk and shear moduli were directly calculated from the velocities of P- and S-waves as $K=\rho[V_{\mathrm{P}}^2-(4/3)V_{\mathrm{S}}^2]$ and $G=\rho V_{\mathrm{S}}^2$, respectively, where $\rho$ is the known rock density.\\

  The pressure dependence of static bulk modulus (and its inverse, compressibility) is obtained from the local slope of volumetric strain \textit{vs.} confining pressure curve (Figure \ref{fig:staticdynamic}b). For many rocks, it is reasonable to assume that the rock compressibility, $K^{-1}$, decays exponentially with pressure as
  \begin{equation}
    \label{eq:CP}
    K^{-1}=K_{\infty}^{-1}+(K_0^{-1}-K_{\infty}^{-1}) e^{-P_{\textrm{c}}/\hat{P}},
  \end{equation}
  where $K_0^{-1}$ and $K_{\infty}^{-1}$ refer to zero- and high-pressure compressibilities, respectively \citep{RWZ1991}. Using such a function is useful to extract values of rock compressibility as well as the characteristic ``crack closure pressure'', $\hat{P}$ (see section \ref{subsubsec:discuss_data}). By integrating equation \ref{eq:CP}, the pressure-dependence of volumetric strain is given by
  \begin{equation}
    \label{eq:epsP}
    \epsilon_{\mathrm{v}}=K_{\infty}^{-1}P_{\textrm{c}}-(K_0^{-1}-K_{\infty}^{-1}) \hat{P} e^{-P_{\textrm{c}}/\hat{P}},
  \end{equation}
  which was used to fit the data in Figure \ref{fig:staticdynamic}b. Pressure-dependent static bulk modulus was then calculated using fitting parameters in equation \ref{eq:CP}.

  An error of 3$\%$ (5$\%$) on $V_{\mathrm{P}}$ ($V_{\mathrm{S}}$) measurements propagates onto a error of 15$\%$ (10$\%$) on $K$ ($G$). An upper bound on the error in static measurements is estimated from the difference in local strain gauge measurements from one pair of strain gauges to another, and is $<$15$\%$ on both static bulk and shear moduli. An acceptable agreement is found between static and dynamic moduli, especially at high pressures. Poisson's ratio is $\approx$ 0.26 and $\approx$ 0.23 for static and dynamic measurements, respectively, which is consistent with the Voigt-Reuss-Hill average from single-crystal, Brillouin scattering data (0.26) \citep{Bezacieretal2010}) and slightly lower than the value obtained from ultrasonic velocity data on bulk antigorite serpentinite (0.29) \citep{Christensen1978}). Another important result of Figure \ref{fig:staticdynamic} is that the pressure-dependence of moduli from mechanical and acoustic measurements are very different. Dynamic moduli markedly increase over the 0--150~\si{\mega\pascal} available experimental range, whereas any measurable variation of static moduli with hydrostatic pressure occurs below \SI{50}{\mega\pascal}.

  \begin{figure}[t]
    \centering
    \includegraphics[width=0.42\textwidth]{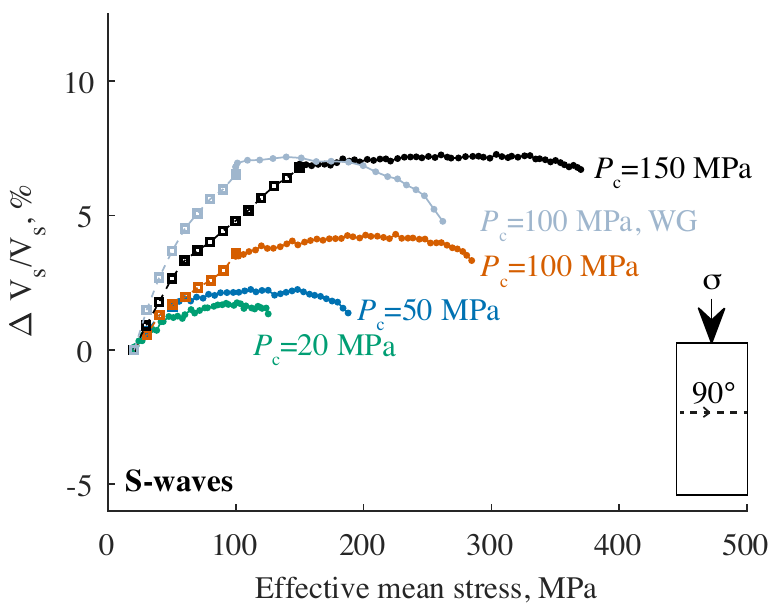}
    \caption{S-wave velocity data (relative change) for all direct loading tests until rock failure, on antigorite at various confining pressures $P_{\textrm{c}}$, and Westerly granite (WG) at $P_{\textrm{c}}$~=~\SI{100}{\mega\pascal} (see Table \ref{tab:samples}), at \ang{90} with respect to the compression axis. Open squares: hydrostatic loading; full circles: axial loading. Note that S-wave velocity data are only available to about 80$\%$ of failure stress in deformation test on Westerly granite, due a technical issue.} 
    \label{fig:velocS_direct}
  \end{figure}

  \begin{figure}[p]
    \centering
    \includegraphics[width=\textwidth]{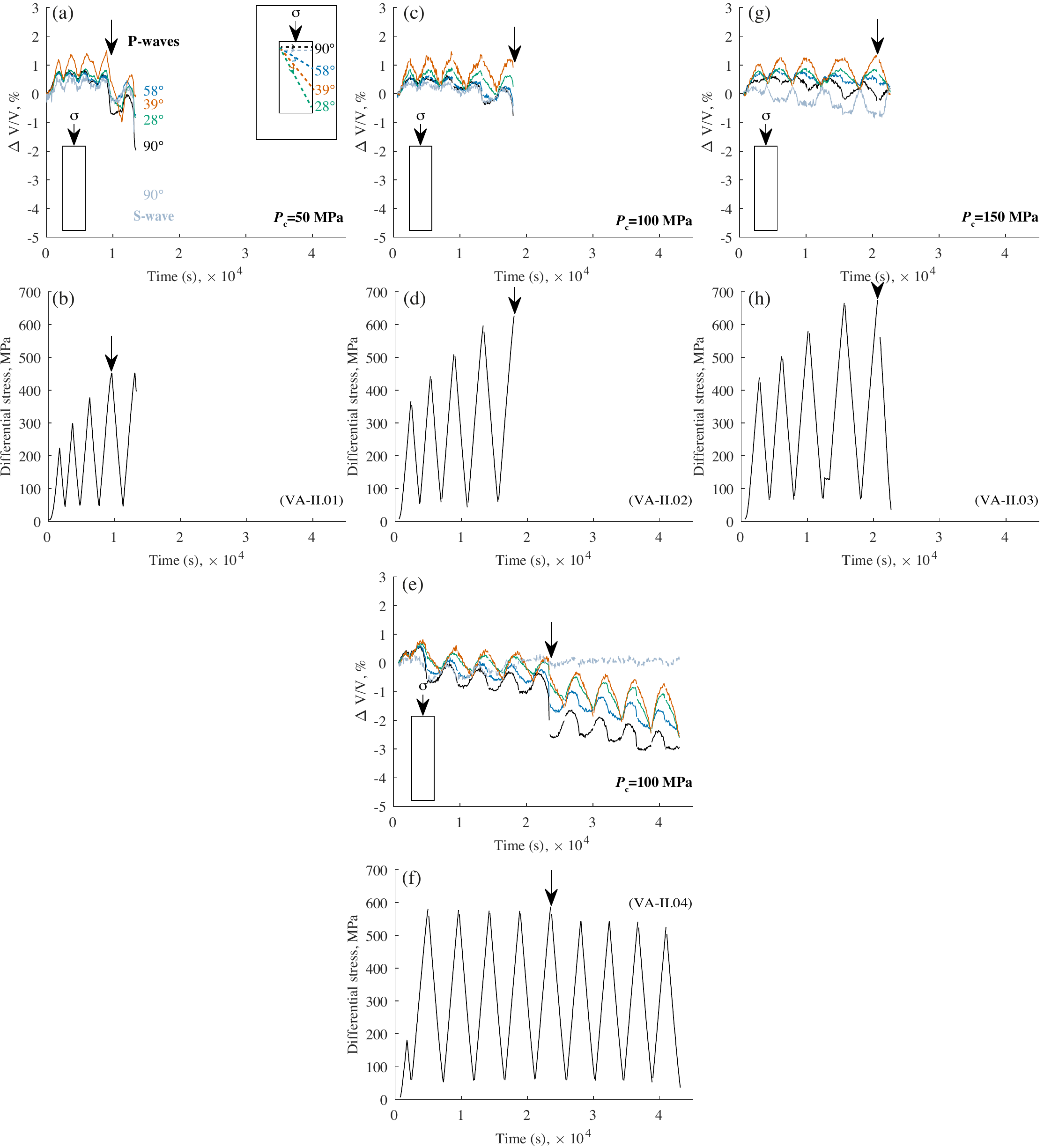}
    \caption{P- and S-wave velocities (relative changes) and differential stress for comparison as functions of time for all cyclic loading tests on the \emph{isotropic antigorite} samples (see Table \ref{tab:samples}) at various confining pressures:  (a,b) $P_{\textrm{c}}$~=~\SI{50}{\mega\pascal}; (c,d and e,f) $P_{\textrm{c}}$~=~\SI{100}{\mega\pascal}; (g,h) $P_{\textrm{c}}$~=~\SI{150}{\mega\pascal}. P-wave velocities were measured at four angles with respect to direction of axial compression: \ang{90}, \ang{58}, \ang{39} and \ang{28}. S-wave velocities were measured at \ang{90} with respect to direction of axial compression. Arrows on curves indicate \emph{quasi-static rock failure} (see section \ref{subsubsec:exptriax} for details).} 
    \label{fig:veloc_cyclicII}
  \end{figure}

      \begin{figure*}
    \centering
    \includegraphics[width=\textwidth]{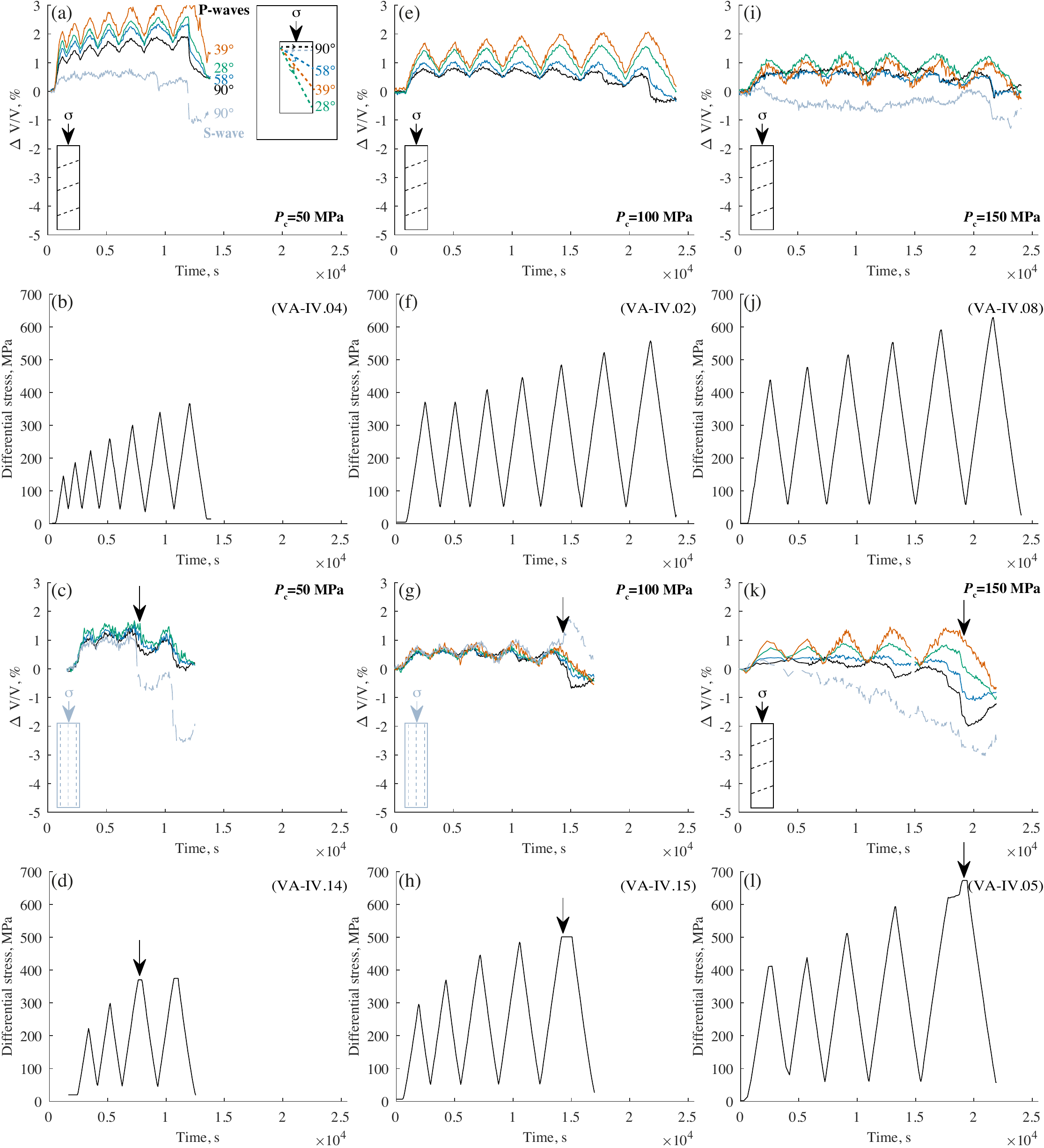}
    \caption{P- and S-wave velocities (relative changes) and differential stress for comparison as functions of time for all cyclic loading tests on the \emph{anisotropic antigorite} samples (see Table \ref{tab:samples}) at various confining pressures: (a,b and c,d) $P_{\textrm{c}}$~=~\SI{50}{\mega\pascal}; (e,f and g,h) $P_{\textrm{c}}$~=~\SI{100}{\mega\pascal}; (i,j and k,l) $P_{\textrm{c}}$~=~\SI{150}{\mega\pascal}. P-wave velocities were measured at four angles with respect to direction of axial compression: \ang{90}, \ang{58}, \ang{39} and \ang{28}. S-wave velocities were measured at \ang{90} with respect to direction of axial compression. Arrows on curves indicate \emph{quasi-static rock failure} (see section \ref{subsubsec:exptriax} for details).} 
    \label{fig:veloc_cyclicIV}
  \end{figure*}

  \subsection{Microstructural Observations in Antigorite Specimens Recovered after Failure}
  \label{subsec:res_micro}
  Representative microstructures in antigorite specimens recovered after rock failure, at confining pressures of 50, 100 and 150~\si{\mega\pascal}, are presented in Figure \ref{fig:micro_deformed}. Additional microstructural observations are reported in Figure \ref{fig:micro_deformed_extra}. \clearpage\par\noindent In all failed specimens, it is first observed that the fault forms an angle of about \ang{30} to the direction of axial compression (\textit{e.g.}, Figures \ref{fig:micro_deformed}a,b,c). In the anisotropic specimens, the foliation does not appear to control the location and orientation of experimental fractures. This is the case for samples in which the foliation is at \ang{70} to the compression axis (Figures \ref{fig:micro_deformed}a,c) and parallel to the compression axis (Figure \ref{fig:micro_deformed_extra}a). Fractures can locally follow a magnesite- and magnetite-rich vein in some portions of the sample (Figure \ref{fig:micro_deformed}c), but also be found in antigorite parallel to a vein (Figure \ref{fig:micro_deformed}a) and pervasively in pure antigorite (Figure \ref{fig:micro_deformed_extra}a). Fractures traverse the magnesite and magnesite grains (Figure \ref{fig:micro_deformed}b), and show no particular deviation around these grains, or veins. These observations indicate that, in all antigorite samples, experimental fractures at sample scale are controlled by the orientation to the axis of compression.

  \begin{figure*}[b]
    \centering
    \includegraphics[width=0.82\textwidth]{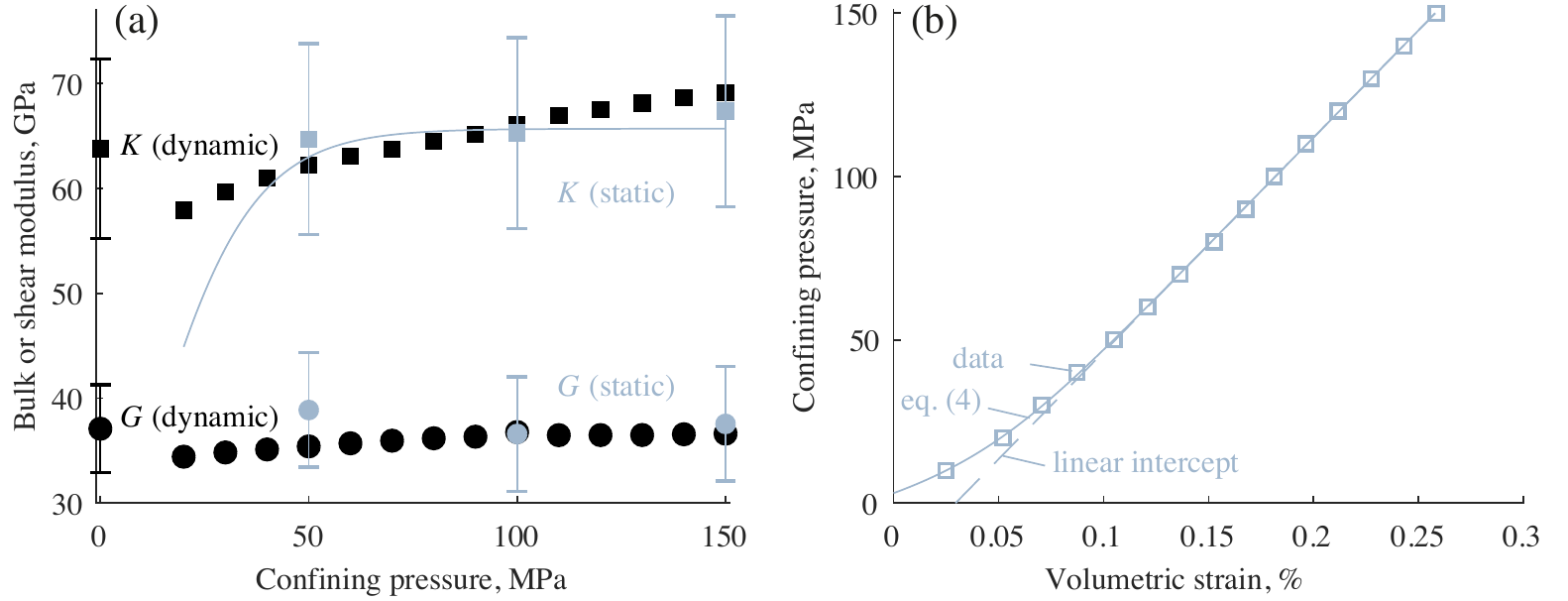}
    \caption{a) Bulk (squares and grey curve) and shear (circles) moduli from static (grey) and dynamic (black) measurements. Static moduli are obtained from the elastic portion of stress-strain curves under axial loading, at three given confining pressures: $P_{\textrm{c}}$~=~\SI{50}{\mega\pascal} (``isotropic'' antigorite specimen, VA--II.1), $P_{\textrm{c}}$~=~\SI{100}{\mega\pascal} (VA--II.2) and $P_{\textrm{c}}$~=~\SI{150}{\mega\pascal} (VA--II.3). The pressure dependence of static bulk modulus (grey curve) is obtained from fitting volumetric strain-confining pressure data (antigorite specimen VA--II.3) -- see section \ref{subsec:res_staticdynamic} for details. Dynamic moduli are calculated from P- and S-wave velocity measurements at increasing confining pressures up to \SI{150}{\mega\pascal} on one antigorite specimen (VA--II.3). Zero-pressure measurements were taken on the ``isotropic'' antigorite block (see section \ref{subsec:spec}). The absolute uncertainty on dynamic bulk and shear moduli is obtained from propagating uncertainties in P- and S-wave velocity measurements; however, under pressure, the relative precision between dynamic measurements is very high (see section \ref{subsubsec:expveloc} for details). b) Volumetric strain \textit{vs.} confining pressure data (squares) for hydrostatic loading on one ``isotropic'' antigorite specimen (VA--II.3). Solid curve: exponential fit (eq. \ref{eq:epsP}); dashed line: linear fit of high pressure data (see section \ref{subsec:res_staticdynamic} for details). Fitting parameters in equation \ref{eq:epsP}: $K_{\infty}$~=~\SI{65.8}{\giga\pascal}; $K_0^{-1}$~=~\SI{20.1}{\per\pascal}; $\hat{P}$~=~\SI{0.015}{\giga\pascal}.} 
    \label{fig:staticdynamic}
  \end{figure*}

  A striking observation is that, except in a region extending typically about 100--200~\si{\micro\metre} on each side of the fault zone, microstructures in fractured antigorite specimens are indistinguishable from the starting rock material (Figures \ref{fig:micro_deformed}a,b,c and \ref{fig:micro_deformed_extra}a), as previously reported by \citet{Escartinetal1997}. The extent of the damage zone around the fault is very similar in antigorite specimens recovered after direct failure (\textit{i.e.}, Figures \ref{fig:micro_deformed}a,b) and quasi-static failure (\textit{i.e.}, Figures \ref{fig:micro_deformed}c and \ref{fig:micro_deformed_extra}a). Damage thus appears to be very localised, and this is the case at all confining pressures. Such damage is formed by elongated ``shear'' microcracks of variable lengths ranging from sub-micron to 10s \si{\micro\metre}, with very small apertures and located along the cleavage planes of antigorite (Figure \ref{fig:micro_deformed}d,e,f), as previously reported by \citet{Escartinetal1997} in a similar confining pressure range. At the grain scale, the orientation of such microcracks seems to be highly controlled by the orientation of the cleavage planes, rather than by the angle to the compression axis (Figure \ref{fig:micro_deformed}f and \ref{fig:micro_deformed_extra}b). 

  Additional microstructural observations, albeit not pervasive across the samples, are reported in Figure \ref{fig:micro_deformed_extra}. Thin films with foam-like matrix ($\sim$\SI{5}{\micro\metre} wide) are found pervasively along the fault plane in one antigorite sample, recovered after direct (\textit{i.e.}, abrupt) failure at $P_{\textrm{c}}$~=~100~\si{\mega\pascal} (Figures \ref{fig:micro_deformed}e and \ref{fig:micro_deformed_extra}c), suggestive of the presence of melt during failure. Similar observations have previously been made in \citet{Brantutetal2016} in a dynamic rupture experiment on a saw-cut antigorite serpentinite at $P_{\textrm{c}}$~=~95~\si{\mega\pascal}. Finally, although rare, kink bands in antigorite grains have been observed within the damage zone in some specimens (\textit{e.g.}, Figure \ref{fig:micro_deformed_extra}d). These additional microstructural features, however, are only observed on or very near the fault plane. This suggests that the formation of kink bands and the possible presence of melt are both likely to be related to the large strains and high slip rate associated with the syn- or post-failure process.

  Overall, microstructural observations indicate that brittle deformation in antigorite is accomodated by very localised shear-dominated microcracking forming preferentially along cleavage planes of antigorite grains.

  \section{Discussion}
  \label{sec:discussion}

  \subsection{Elasticity and Crack Closure in Antigorite}
  \label{subsec:discuss_crackclosure}
  \subsubsection{Dependence of Velocities upon Hydrostatic Pressure in Antigorite: Comparison with Existing Data, and Interpretation in terms of Microcrack Closure}
  \label{subsubsec:discuss_data}

  A comparison of all available ultrasonic P- and S-wave velocity data on polycrystalline antigorite-rich serpentinite, as functions of confining pressure, is given in Figure \ref{fig:ARD}a. Corresponding values of Poisson's ratio, $\nu$, which is calculated from the ratio of P- and S-wave velocities as $\nu=1/2[(V_{\mathrm{P}}/V_{\mathrm{S}})^2-2]/[(V_{\mathrm{P}}/V_{\mathrm{S}})^2-1]$, are shown in Figure \ref{fig:ARD}b. As previously stated, velocities are strongly dependent on serpentine mineralogy and the presence of accessory phases; natural serpentinites also display a strongly variable degree of anisotropy. Hence, for sensible comparison, the two criteria for selection of previously published data in Figure \ref{fig:ARD} are that rock specimens are antigorite-rich ($>80\%$) and \emph{nearly isotropic}. For this latter reason, data of \citet{Kern1993}, \citet{Kernetal1997} and \citet{Watanabeetal2007} are discarded, as the focus of these studies was mostly on velocity anisotropy. Accordingly, for this study, the results shown in Figure \ref{fig:ARD} are for the ``isotropic'' antigorite sample subjected to the highest confining pressure (\SI{150}{\mega\pascal}). All sample details are given in the caption of Figure \ref{fig:ARD}.

  \begin{figure*}
    \centering
    \includegraphics[width=0.85\textwidth]{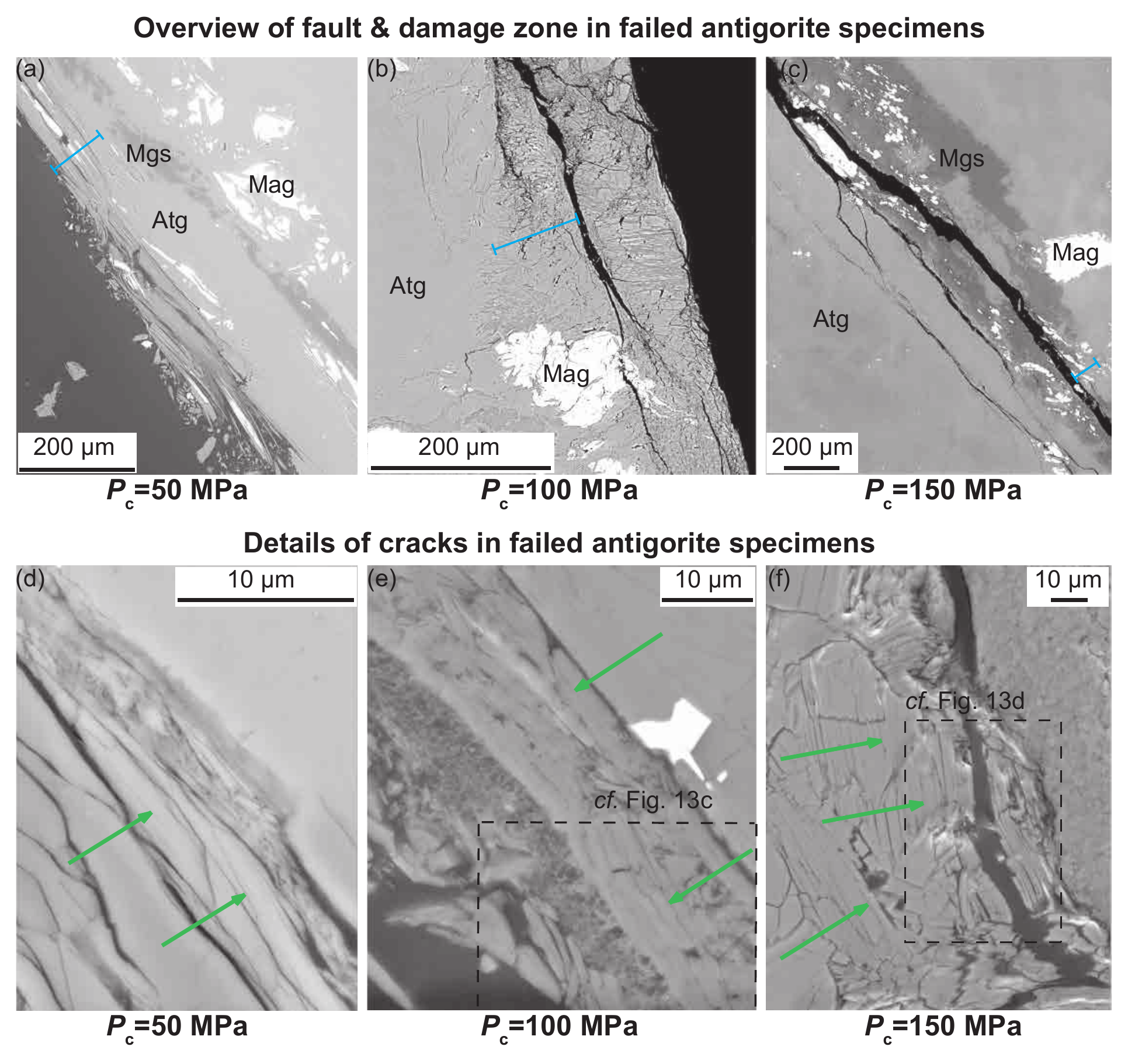}
    \caption{Backscattered-electron images of representative microstructures in antigorite specimens recovered after axial deformation and rock failure, at three different confining pressures $P_{\textrm{c}}$. The fault strike is normal to the plane formed by the thin-sections, and the direction of axial compression is \emph{vertical} in all images. All images were acquired using Oxford Instruments AZtec software on an FEI Quanta 650 field-emission gun scanning-electron microscope in the Department of Earth Sciences at the University of Oxford. a) antigorite specimen after failure at $P_{\textrm{c}}$~=~\SI{50}{\mega\pascal} (VA--IV.03); b) antigorite specimen after failure at $P_{\textrm{c}}$~=~\SI{100}{\mega\pascal} (VA--II.02); c) antigorite specimen after quasi-static failure at $P_{\textrm{c}}$~=~\SI{150}{\mega\pascal} (VA--IV.05). In a, b and c, the blue line is a \SI{100}{\micro\metre} segment to allow easier comparison between antigorite specimens of the extent of the damage zone around the fault zone. d) antigorite specimen after failure at $P_{\textrm{c}}$~=~\SI{50}{\mega\pascal} (VA--IV.03); e) antigorite specimen after failure at $P_{\textrm{c}}$~=~\SI{100}{\mega\pascal} (VA--IV.01); f) antigorite specimen after quasi-static failure at $P_{\textrm{c}}$~=~\SI{150}{\mega\pascal} (VA--IV.05). The green arrows in d, e and f indicate deformation-induced shear cracks following cleavage planes of antigorite grains.} 
    \label{fig:micro_deformed}
  \end{figure*}

  This study provides new measurements of velocities in antigorite in a finely sampled, ``low-pressure'' experimental pressure window (Figure \ref{fig:ARD}a). Both P- and S-wave measurements of this study, as well as derived values of Poisson's ratio, are in very good agreement with existing data, within experimental error. The ``high-temperature'' polytype antigorite is a high-velocity, ``low'' Poisson's ratio $\sim$ 0.25--0.28 (and, equivalently, ``low'' $V_{\mathrm{P}}/V_{\mathrm{S}}$ ratio $\sim$ 1.73--1.81) end-member among serpentines. Seismologically observed values of Poisson's ratios ($\sim$ 0.30--0.35) that are high compared to those typical of mantle rocks ($\sim$ 0.25) are often interpreted by seismologists to indicate the presence of serpentine and quantify the degree of serpentinization in subduction zones (\textit{e.g.}, \citet{HyndmanPeacock2003,CarlsonMiller2003,DeShonSchwartz2004}). However, experimental measurements of velocities on isotropic serpentinized peridotites \citep{Christensen1966,Horenetal1996,Jietal2013,Shaoetal2014} show that such high values of Poisson's ratio are most probably indicative of lizardite/chrysotile serpentinites, rather than antigorite serpentinites. A high Poisson's ratio is therefore not a clear indicator of the presence of antigorite serpentinites \citep{Reynard2013,Jietal2013}. Similarly, the decrease of velocity with increasing degree of serpentinization between peridotite and antigorite-serpentinite is much less than with between peridotite and lizardite or chrysotile \citep{Christensen2004,Jietal2013}. Velocities in antigorite may be higher than some crustal rocks at similar pressures, such as granite (Figure \ref{fig:ARD}).

  It is well known that velocities in rocks are very sensititive to the presence of open microcracks, and that the increase of ultrasonic velocities with increasing hydrostatic pressure is accordingly associated with progressive closure of such microcracks (\textit{e.g.}, \citet{NurSimmons1969a,Kernetal1997,DavidRWZ2012}). Such interpretation is given to explain the noticeable increase of both P- and S-wave velocities with increasing hydrostatic pressure in our serpentinite samples (and granite sample). The absence of a velocity ``plateau'' (at $P_{\textrm{c}}$~=~\SI{150}{\mega\pascal}) indicates ongoing crack closure, \textit{i.e.}, the presence of a fraction of microcracks remaining open at this pressure. The shape of the velocity \textit{vs.} confining pressure curve suggests that crack closure is still ongoing (albeit probably marginal) at several hundreds of \si{\mega\pascal}s in antigorite.

  Nevertheless, the ``high-pressure'' velocities of the crack-free material can be quantitatively estimated recalling that the rock compressibility can be assumed to decay exponentially with pressure (equation \ref{eq:CP}). By analogy, the same empirical pressure dependence can be assumed for the shear compliance, $G^{-1}$, where ($G_0^{-1}$,$G_{\infty}^{-1}$) are the zero- and high-pressure shear compliance, respectively \citep{DavidRWZ2012}. For antigorite, the ($K^{-1}$,$G^{-1}$) data, which were directly calculated from ($V_{\mathrm{P}}$,$V_{\mathrm{S}}$) data, are jointly fitted by
  \begin{eqnarray}
    K^{-1}&=&0.0137+(0.0181-0.0137)e^{-P_{\textrm{c}}/0.0876} \label{eq:fitCveloc} \ \textrm{(in \si{\per\giga\pascal})},\\
    G^{-1}&=&0.0266+(0.0295-0.0266)e^{-P_{\textrm{c}}/0.0876} \label{eq:fitSveloc} \ \textrm{(in \si{\per\giga\pascal})},
  \end{eqnarray}
  which yields ``high-pressure'' bulk and shear moduli $K_{\infty}$~=~\SI{72.9}{\giga\pascal} and $G_{\infty}$~=~\SI{37.6}{\giga\pascal}, and Poisson's ratio $\nu_{\infty}=(3K_{\infty}-2G_{\infty})/(6K_{\infty}+2G_{\infty})$~=~0.28. The corresponding high-pressure velocities are $V_{\mathrm{P,}\infty}$~=~\SI{6.78}{\kilo\meter\per\second} and $V_{\mathrm{S,}\infty}$~=~\SI{3.75}{\kilo\meter\per\second}. The latter values are very close to measurements of \citet{Christensen1978} at the highest available pressure ($V_{\mathrm{P}}$~=~\SI{6.69}{\kilo\meter\per\second}, $V_{\mathrm{S}}$~=~\SI{3.62}{\kilo\meter\per\second} and $\nu$~=~0.29, at $P_{\textrm{c}}$~=~\SI{1}{\giga\pascal}). High-pressure elastic constants also show excellent agreement with Voigt-Reuss-Hill average of single-crystal elastic constants from \citet{Bezacieretal2013} ($K$~=~\SI{75.3}{\giga\pascal}, $G$~=~\SI{40.1}{\giga\pascal} and $\nu$~=~0.27, at \SI{2}{\giga\pascal}).

  Both P- and S-wave velocities, as well as Poisson's ratio, seem to be noticeably more pressure dependent than previously available data on antigorite in the 20--150~\si{\mega\pascal} pressure range (Figure \ref{fig:ARD}). A similar trend is clearly observed for all antigorite specimens of this study (see, for instance, Figure \ref{fig:velocP_direct} and \ref{fig:velocS_direct}). In particular, the denominator in the exponential term of equations \ref{eq:fitCveloc} and \ref{eq:fitSveloc} is a direct estimate of the ``characteristic pressure for crack-closure'', $\hat{P}$ (equation \ref{eq:CP}). For antigorite, and based on dynamic measurements, $\hat{P}$ is thus about \SI{88}{\mega\pascal}. This value can be compared to other experimental datasets, or pressure-dependence of other physical properties, such as mechanical or ``static'' data (see section \ref{subsubsec:discuss_static}). \citet{Jietal2013} use an empirical fit of pressure-dependent velocity data that incorporates an additional term, linear in pressure, as $V(P_{\textrm{c}})=V_0+DP-B_0e^{-kP_{\textrm{c}}}$, where $V$ is P- or S-wave velocity, and ($V_0$,$D$,$B_0$,$k$) are fitting parameters (such fit is probably more adequate for experiments conducted in a higher pressure than in this study). On an essentially isotropic, 98$\%$ antigorite sample (WZG8), conversion of their coefficient $k$ into $\hat{P}$ for P-wave velocity data gives $\hat{P}$~=~\SI{34}{\mega\pascal}. As for other previous studies shown in Figure \ref{fig:ARD}, such differences in pressure-dependent behaviour could be explained simply by rock variability  in terms of the existing population of microcracks in the starting material, or by the spacing of experimental measurements in the low confining pressure range.

  As pointed out by \citet{Watanabeetal2007}, the increase of S-wave velocity with pressure is less than for P-wave velocity (Figures \ref{fig:velocraw_direct}, \ref{fig:velocP_direct} and \ref{fig:velocS_direct}). Correspondingly, the increase of bulk modulus of the rock with pressure is greater than for shear modulus, and Poisson's ratio increases slightly with pressure (as does $V_{\mathrm{P}}/V_{\mathrm{S}}$ ratio). Such pressure dependence of Poisson's ratio (which is also observed in granite) is entirely consistent with what is expected for three-dimensional micromechanical models deriving elastic properties of dry isotropic solids containing thin spheroidal cracks, using ``effective medium schemes'' (\textit{e.g.}, \citep{OB1974,Berrymanetal2002,DavidRWZ2011}). According to such calculations, the addition of cracks always decreases Poisson's ratio and drives it to zero (in the high-concentration limit), independently of a solid's Poisson's ratio \citep{DavidRWZ2011}. The rather moderate dependence of Poisson's ratio on pressure compared to other rocks (\textit{e.g.}, sandstones \citep{DavidRWZ2012}) is explained by the overall modest amount of cracks in the antigorite (and granite) samples. The ``differential effective medium scheme'' (\textit{e.g.}, \citet{RWZ1984}) is used to invert crack aspect ratio distributions in the following section.

        \begin{figure*}[b]
    \centering
    \includegraphics[width=0.85\textwidth]{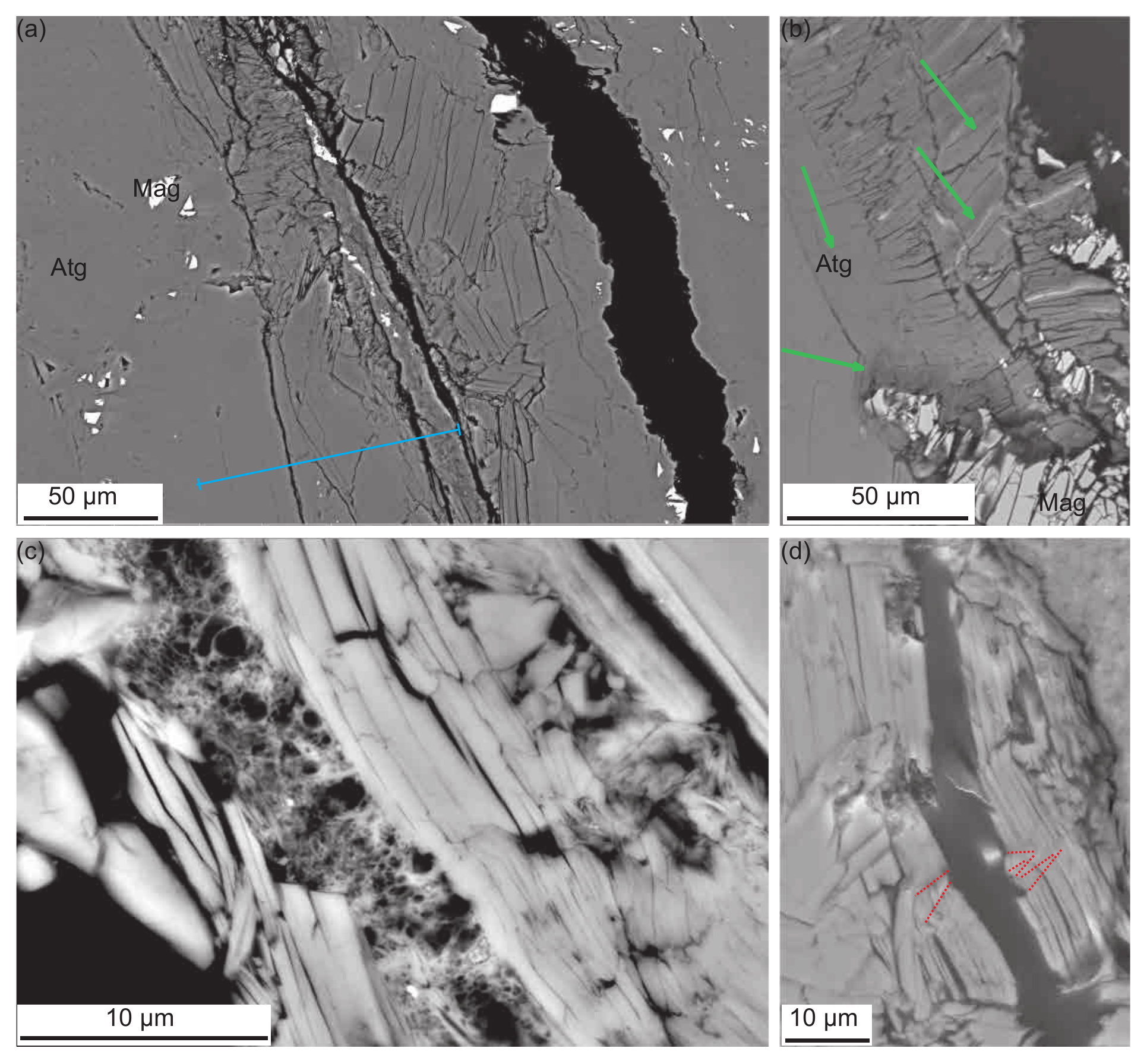}
    \caption{Additional microstructural observations in antigorite specimens recovered after axial deformation and rock failure. The fault strike is normal to the plane formed by the thin-sections, and direction of axial compression is \emph{vertical} in all images. For details of image acquisition, see caption of Figure \ref{fig:micro_deformed}. a) antigorite specimen after quasi-static failure at $P_{\textrm{c}}$~=~\SI{100}{\mega\pascal} (VA--IV.15); the blue line is a \SI{100}{\micro\metre} segment to allow easier comparison between antigorite specimens of the extent of the damage zone around the fault zone. b) antigorite specimen after quasi-static failure at confining pressure $P_{\textrm{c}}$~=~\SI{150}{\mega\pascal} (VA--IV.05). The green arrows indicate deformation-induced shear cracks following cleavage planes of antigorite grains. c) antigorite specimen after failure at confining pressure $P_{\textrm{c}}$~=~\SI{100}{\mega\pascal} (VA--IV.01) (location denoted in Figure \ref{fig:micro_deformed}e). d) antigorite specimen after quasi-static failure at confining pressure $P_{\textrm{c}}$~=~\SI{150}{\mega\pascal} (VA--IV.05) (location denoted in Figure \ref{fig:micro_deformed}f). The red dashed lines highlight kink bands. } 
    \label{fig:micro_deformed_extra}
  \end{figure*}

  \subsubsection{Inversion of Microcrack Aspect Ratio Distribution from the Dependence of Velocities upon Hydrostatic Pressure}
  \label{subsubsec:discuss_ARD}
  The pressure dependence of P- and S-wave velocities can be inverted to extract crack aspect ratio distribution on dry isotropic rock \citep{RWZ1991,DavidRWZ2012}. The model of \citet{DavidRWZ2012}, initially developed for and succesfully tested on sandstones containing both pores and cracks, is applied to polycrystalline antigorite and granite samples containing only cracks. The crack-free, non-porous ``matrix'' formed by the minerals is assumed to contain a population of spheroidal cracks, having an aspect ratio $\alpha$ (defined as the ratio of crack aperture over crack length). Such population of cracks is well described by the mean volumetric crack density parameter defined as $\Gamma=N_{\textrm{V}}\langle l^3 \rangle$, where $N_{\textrm{V}}$ is the number of cracks per volume, $l$ is crack length, and the angle brackets indicate an arithmetic average. The differential effective medium scheme is used here to calculate the impact of a given crack population on elastic moduli. As the required pressure to close a thin spheroidal crack is proportional to its aspect ratio \citep{Walsh1965}, the non-linear elastic behaviour during hydrostatic compression can be accounted for by assuming that the rock contains an (exponential) distribution of aspect ratios \citep{RWZ1991}.

\begin{figure*}
    \centering
    \includegraphics[width=0.82\textwidth]{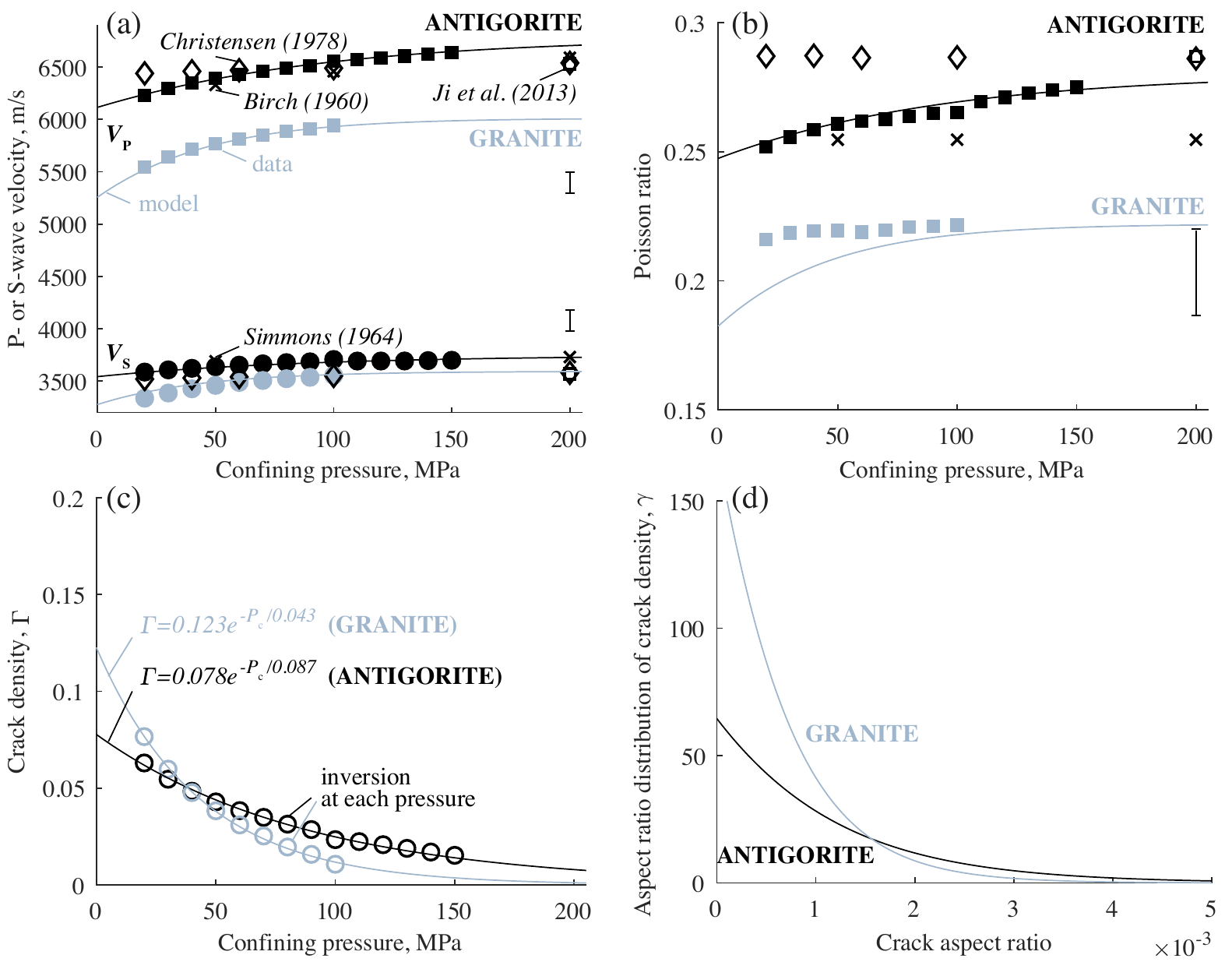}
    \caption{Comparison of a) P-, S-wave velocity and b) Poisson's ratio data (symbols in a,b), with previous experimental studies on antigorite-rich, nearly isotropic serpentinites, and inversion of c) crack density and d) aspect ratio distribution, in Vermont antigorite (specimen VA--II.3, black curves) and Westerly granite (WG2, grey curves), based on the spheroidal crack closure model of \citet{DavidRWZ2012} (see text for details). a,b) Full squares (circles): P- (S-)wave velocity data (this study; note that, at given pressure, the P-wave velocity is independent of the measurement direction). Crosses: Ludlow antigorite serpentinite (86$\%$ antigorite) \citep{Birch1960,Simmons1964}; open diamonds: Stonyford serpentinite (2) (95$\%$ antigorite) \citep{Christensen1978}; open squares: equations (7) and (9) of \citet{Jietal2013} fitted to a suite of 17 antigorite serpentinite samples and taken for pure antigorite at $P_{\textrm{c}}$~=~\SI{200}{\mega\pascal}. For details on errorbars in velocity measurements in a and b, see section \ref{subsubsec:expveloc}. c) Inverted cumulative crack density, $\Gamma$, at each pressure (open circles), and exponential empirical fits ($P_{\textrm{c}}$ in \si{\giga\pascal}); d) Inferred aspect ratio distributions of crack density, $\gamma$. } 
    \label{fig:ARD}
  \end{figure*}

  The model of \citet{DavidRWZ2012} was applied to the data obtained on the isotropic antigorite and granite samples, and implemented as follows. The high-pressure, ``crack-free'' elastic properties of the minerals are first obtained by jointly fitting the pressure-dependence of compressibility and shear compliance data (equations \ref{eq:fitCveloc} and \ref{eq:fitSveloc}; fitting parameters for granite are: $K_0$~=~\SI{38.5}{\giga\pascal}; $K_{\infty}$~=~\SI{50.0}{\giga\pascal}; $G_0$~=~\SI{27.2}{\giga\pascal}; $G_{\infty}$~=~\SI{34.1}{\giga\pascal}; $\hat{P}$~=~\SI{0.043}{\giga\pascal}). At each pressure, a given crack density (Figure \ref{fig:ARD}c) is then inverted from moduli deficits relative to the ``high-pressure'' elastic moduli of minerals ($K_{\infty}$,$G_{\infty}$). Values of $\Gamma (P_{\textrm{c}})$ thus determined are fitted by an exponential function $\Gamma=\Gamma_0 e^{-P_{\textrm{c}}/\hat{P}}$ (Figure \ref{fig:ARD}c), using a fixed parameter $\hat{P}$ already determined from empirical fits of pressure-dependent compressibility and shear compliance data, and where $\Gamma_0$ is thus the estimated zero-pressure crack density. At this stage, the micromechanical model is able to describe the pressure-dependence of elastic constants and velocities (Figure \ref{fig:ARD}a,b) by simply incorporating the exponentially decaying, pressure-dependent crack density into the differential scheme equations \citep{DavidRWZ2012}. Finally, by recalling that cracks close at a pressure proportional to their aspect ratio $\alpha$, an adequate change of variables from pressure to aspect ratio directly converts $\Gamma (P_{\textrm{c}})$ into $\Gamma(\alpha)$, the \emph{aspect ratio distribution function} of the crack density (in the cumulative sense) \citep{DavidRWZ2012}. Accordingly, the aspect ratio distribution function of crack density shown in Figure \ref{fig:ARD}d is simply $\gamma(\alpha)=d\Gamma / d\alpha$.

  The model provides a very good fit to the pressure-dependence of both P- and S-wave velocity data, for both antigorite and granite (Figure \ref{fig:ARD}a), as well as Poisson's ratio for antigorite (Figure \ref{fig:ARD}b). This goodness of fit is in part due to the fact that compressibility and shear compliance data, as well as the values of crack density inverted by the model, are all well described by functions that decay exponentially with pressure. Note that the relatively poor fit of Poisson's ratio for granite at low confining pressures originates from the high sensitivity of Poisson's ratio to the small misfits in P- and S-wave velocities, a reason for which the $V_\textrm{P}$/$V_\textrm{S}$ ratio is often preferred over the Poisson's ratio in seismology \citep{Thomsen1990}. The zero-pressure crack density in antigorite is less than that for granite (Figure \ref{fig:ARD}c), but its distribution of crack aspect ratios is broader, with a mean aspect ratio about 2$\times10^{-3}$. Accordingly, the characteristic pressure for crack closure in antigorite ($\hat{P}$~=~\SI{88}{\mega\pascal}) is twice that of granite (\SI{43}{\mega\pascal}), and the indicative pressure at which crack closure is essentially complete is about 400 and 200~\si{\mega\pascal} in antigorite and granite, respectively. Another useful outcome of the effective medium modelling is that the total crack porosity can be directly estimated from the crack aspect ratio distribution. By recalling that the crack porosity is related to crack density as $c(\alpha)=4\pi\alpha\gamma(\alpha) / 3$ \citep{RWZ1991}, the total crack porosity is simply the cumulative integral of the crack porosity distribution function $c(\alpha)$. For antigorite, the calculated crack porosity is 0.036$\%$, which agree with Helium pycnometry measurements used in sample characterisation (see section \ref{subsec:spec}).

  \subsubsection{Comparison of Pressure Dependence of Dynamic and Static Moduli in Antigorite}
  \label{subsubsec:discuss_static}
  Static and dynamic properties are known to be both sensitive to the closure of microcracks (and, more generally, to the \emph{presence} of open microcracks) in rocks \citep{PatersonWong}. The crack porosity can also be estimated from mechanical (static) data, by reading the zero-pressure intercept of the linear portion of the volumetric strain \textit{vs.} hydrostatic pressure curve \citep{Vajdovaetal2004}. For antigorite, the indicative crack porosity calculated from strain gauge data is about 0.03$\%$ (Figure \ref{fig:staticdynamic}b), which compares very well to crack porosity calculated from velocity data (0.036$\%$, see previous section). In addition, both static and dynamic measurements yield consistent values of elastic moduli, especially as pressure increases. However, the pressure dependence of elastic moduli derived from both physical measurements is noticeably different. Fitting compressibility-pressure mechanical data (see section \ref{subsec:res_staticdynamic}) yields a characteristic pressure for crack closure $\hat{P}\approx$\SI{15}{\mega\pascal} which is much less than that inferred from velocity measurements ($\hat{P}\approx$\SI{88}{\mega\pascal}). Values of tangent bulk moduli at low pressures are also much less than bulk moduli inferred from dynamic measurements (Figure \ref{fig:staticdynamic}a). Such differences between static and dynamic moduli are well above experimental error. The wavelength in dynamic measurements is typically in the 5--7~\si{mm} range, which compares with the size of strain gauges ($\sim$\SI{10}{\milli\metre}), and is at least two to three orders of magnitude greater than grain or crack size. Strain gauges provide local measurements, which could be affected by rock heterogeneity, such as foliation veins, or accessory minerals; however, comparison of static and dynamic moduli on the anisotropic antigorite specimen pressurised at \SI{150}{\mega\pascal} yields similar results as in Figure \ref{fig:staticdynamic}. The probable source of the discrepancy has to do with cracks being open at low pressures in antigorite, and that strain gauge measurements are more likely to capture elastic strains associated with the closure of cracks than a travelling pulse potentially bypassing such cracks, as pointed out by a number of experimental studies on polycrystalline rocks \citep{SimmonsBrace1965,King1983}.

  \subsection{Absence of Stress-induced Seismic Anisotropy during Antigorite Brittle Deformation}
  \label{subsec:discuss_aniso}

  The most impactful observation in this series of experiments is the spectacular absence of stress-induced seismic anisotropy during antigorite brittle deformation up to failure, in the entire confining pressure range below \SI{150}{\mega\pascal} ($\sim$\SI{5}{\kilo\metre} depth in Earth's crust), at room temperature. Such untypical behaviour does not seem to have been reported in crystalline rocks before, and is significantly different from observations on granite (\textit{e.g.}, this study; \citep{NurSimmons1969b,Sogaetal1978}) and many other crystalline rocks (see, for instance, \citet{PatersonWong}).

  The evolution of P- and S-wave velocities observed during brittle deformation of the Westerly granite specimen at $P_{\textrm{c}}$~=~\SI{100}{\mega\pascal} (Figures \ref{fig:velocP_direct} and \ref{fig:velocS_direct}) is typical of that of crystalline rocks and results from both closure of existing microcracks and, at sufficiently high differential stress, opening (propagation) of new ``mode I'' microcracks \citep{PatersonWong}. In the early stage of granite deformation, to about 50$\%$ of failure stress, the slight increase in velocities is explained by the elastic closure of existing cracks, \textit{i.e}, of the population of cracks that are still open at $P_{\textrm{c}}$~=~\SI{100}{\mega\pascal} (see above). As the closure of cracks is more favourable for cracks whose orientations are \emph{sub-perpendicular} to the compression axis, wave velocities measurably increase as the direction of wave propagation approaches the compression axis (Figure \ref{fig:velocP_direct}a to d). From about 50$\%$ of granite failure stress, the elastic closure of cracks is then progressively dominated by the inelastic propagation (opening) of axial cracks -- an accelerating process leading to rock failure and causing a large decrease in velocities (up to 15$\%$ for P-waves). As the opening of cracks is more favourable for cracks whose orientation is \emph{sub-parallel to the compression axis}, the decrease in wave velocities is greater (and occurs earlier) in the direction perpendicular to the compression axis than sub-parallel to it (Figure \ref{fig:velocP_direct}a to d). Hence, under axial compression, the opening of axial microcracks and, to a minor extent, the elastic closure of cracks, both give rise to a pronounced stress-induced anisotropy, which is about 10$\%$ for P-waves at failure (Figure \ref{fig:velocP_direct}).

  The quantitative inversion of the pressure-dependence of velocities into crack aspect ratio distribution (see above) demonstrates that Vermont antigorite has a measurable density of \emph{existing} microcracks ($\Gamma$~=~0.08) that is comparable to that of Westerly granite ($\Gamma$~=~0.12), but also suggests that a non-negligible fraction of cracks in antigorite remains open at the confining pressures used in this series of deformation experiments (\textit{e.g.}, about 25$\%$ even at $P_{\textrm{c}}$~=~\SI{150}{\mega\pascal}, Figure \ref{fig:ARD}c). The slight increase of both P and S-wave velocities in antigorite during axial deformation, observed at all confining pressures and preferably in directions sub-parallel to the compression axis (Figure \ref{fig:velocP_direct}), is thus explained by the elastic closure of such cracks. A major difference with granite is that the elastic closure of cracks in antigorite dominates the evolution of velocities almost up to failure. Hence, in antigorite, the absence of significant decrease in velocities and of stress-induced anisotropy even prior to failure convincingly indicates that the process of brittle failure in antigorite is not associated with any notable opening of ``mode I'' microcracks and, more generally, with any pervasive damage. This is observed even at low confining pressure conditions that are, in principle, more favourable to the opening of microcracks \citep{PatersonWong}.

  We can place an upper bound on the density of such ``stress-induced'' microcracks that may arise during deformation of antigorite. The maximum decrease of P-wave velocity (3$\%$) in all experiments in antigorite is observed in the direct failure test at \SI{150}{\mega\pascal}, at \ang{90} to the applied stress. If ($x_1$,$x_2$,$x_3$) denote orthogonal directions, where ($x_3$) is aligned with the rock cylinder axis, in a transversely isotropic medium, the elastic stiffness tensor component $C_{11}$ (the ``P-wave modulus in direction $x_1$'') is directly related to $V_{\textrm{P}}(90^{\circ})$ as
  \begin{equation}
    \label{eq:V11C11}
    C_{11}=\rho V_{\textrm{P}}(90^{\circ})=\rho V_{11}
  \end{equation}
  \citep{SayersKachanov1995}. Similarly to hydrostatic loading, the modulus deficit relative to the uncracked state can be related to a microcrack density. Note that \citet{SayersKachanov1995} consider the case of thin ``penny-shaped'' cracks rather than thin spheroidal cracks, but analytical solutions for both cases, in the case of dry cracks, converge in the limit of small crack aspect ratios. It is reasonable to consider the case of cracks with normals randomly oriented within planes parallel to the $x_1x_2$ plane (\textit{i.e.}, cracks are parallel to the applied stress), which is described in \citet{SayersKachanov1995}. Combining equations (22), (23) and (28) of that paper, and neglecting the contribution of the fourth-rank crack density tensor (an approximation valid for dry rocks), it is found that $C_{11}$, the normalised elastic stiffness in direction $x_1$, can be directly related to the crack density in that direction, $\gamma_{11}$, in close-form as
  \begin{equation}
    \label{eq:C11g11}
    \frac{C_{11}}{C_{11}^0}=1-\bigg[ \frac{32(1-2\nu_0+2\nu_0^2)}{3(2-\nu_0)(1-2\nu_0)} \bigg] \gamma_{11},
  \end{equation}
  by taking a Taylor series expansion for small values of crack density. $\nu_0$ is the Poisson's ratio of the minerals ($\nu_0$~=~0.28) and $C_{11}^0$ is the uncracked ``isotropic'' P-wave modulus ($C_{11}^0$~=~\SI{123.0}{\giga\pascal}), both taken at high pressure (see section \ref{subsubsec:discuss_data}). The velocity-stiffness relation above (\ref{eq:V11C11}) shows that a 3$\%$ variation of $V_{11}$ is equivalent to a 6$\%$ variation of $C_{11}$, which gives $\gamma_{11}$~=~0.007. The added total crack density of cracks opening in a direction parallel to the compression axis is thus $\gamma=2\gamma_{11}$~=~0.014, considering the cracks in the equivalent direction $x_2$. This value is much smaller to that obtained doing the same calculation for Westerly granite at $P_{\textrm{c}}$~=~\SI{100}{\mega\pascal} ($\gamma$~=~0.09) or at lower confining pressures (\textit{e.g.}, \citet{Sogaetal1978}). In antigorite, the amount of ``stress-induced'' \textit{vs.} ``zero-pressure, existing'' cracks is also small compared to granite (20$\%$ and 75$\%$, respectively). Note that the calculations of \citet{SayersKachanov1995} use the non-interactive effective medium scheme, whereas the inversion described in section \ref{subsubsec:discuss_ARD} uses the differential scheme; however, at small crack densities ($<$ 0.1), values of crack density inverted by both effective medium models differ by a negligible amount \citep{RWZ1991}.

  \subsection{Evolution of Velocities during Quasi-static Failure in Antigorite}
  \label{subsec:discuss_velocfailure}

  In addition to the considerations given in previous section, the absence of any significant decrease in wave velocities and of stress-induced anisotropy prior to brittle failure also suggests that any opening of axial microcracks in antigorite would be very localised and/or occurring only immediately prior to failure. Indeed, the ``volumetric'' contribution of such cracks is not necessarily expected to be detectable in the velocity data (as shown) since these data are averages of multiple raypaths at a given angle to the compression axis (see section \ref{subsubsec:expveloc} for geometry of wave velocity sensors and methodology), meaning that when strain becomes localised on a fault, some raypaths intersect the fault, while some others do not. In addition, velocity surveys are collected every minute, so the opening of microcracks would not necessarily be captured if it occurs immediately prior to failure, notably during brittle failure tests.

  \begin{figure}
    \centering
    \includegraphics[width=0.45\textwidth]{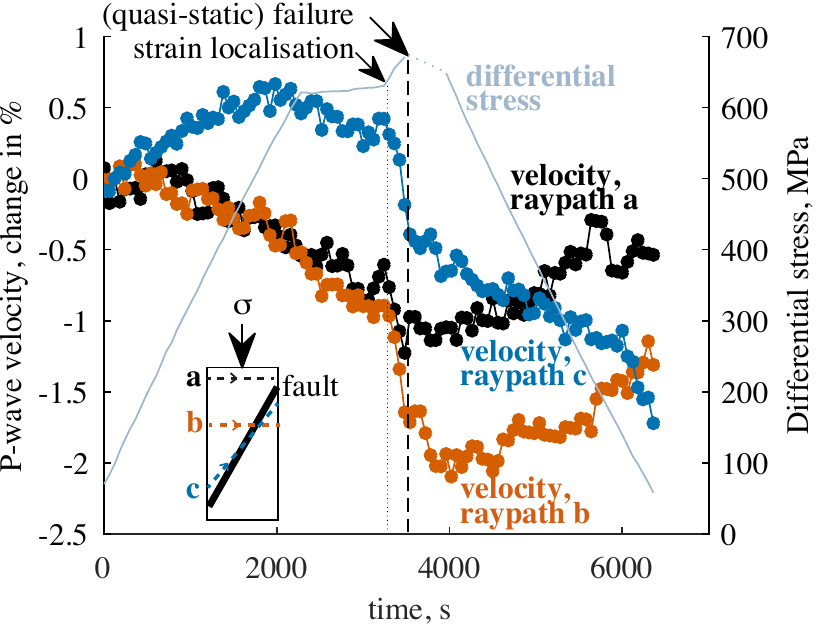}
    \caption{Evolution of P-wave velocities (and differential stress) as function of time during quasi-static failure, at $P_{\textrm{c}}$~=~\SI{150}{\mega\pascal} (specimen VA--IV.05). Three individual raypaths were selected (see section \ref{subsubsec:expveloc}): one raypath off-fault (black dots); one raypath across fault, at an angle $\sim$\ang{60} to the fault (red dots); one raypath across fault and almost parallel to it, at an angle $\sim$\ang{10} to the fault (blue dots).} 
    \label{fig:Vfault}
  \end{figure}

  Hence, in order to examine the extent to which damage is localised and observe the evolution of velocities at the onset of rock localisation and \emph{during} failure, it is attractive to look at the evolution of velocities over \emph{specific} raypaths during quasi-static failure tests. A representative example of P-wave velocity evolution during quasi-static failure, at \SI{150}{\mega\pascal} confining pressure, is shown in Figure \ref{fig:Vfault}. The location of the fault plane with respect to the geometrical arrangement of sensors could be precisely determined on the rock specimen recovered after the experiment, allowing selection of three specific P-wave raypaths: (a) one horizontal raypath out of the fault plane, in the upper part of the rock specimen; (b) one horizontal raypath across the fault plane, forming an angle of approximately \ang{60} to the fault plane; (c) one ``diagonal'' raypath (at \ang{39} to the compression axis), sub-parallel or ``along'' the fault plane (c). Note that the change of slope in the differential stress \textit{vs.} time curve (plotted here for comparison), for a $\sim$\SI{1000}{\second} duration prior to rock failure, corresponds to a change in axial piston advancement rate as part of the strategy to achieve quasi-static failure on the antigorite specimen (see section \ref{subsubsec:exptriax}). With increasing differential stress and before strain becomes localised (see below), the two horizontal P-wave velocities decrease by about 1$\%$, while the sub-vertical P-wave velocity first increases by 0.5$\%$ and then starts decreasing at about \SI{600}{\mega\pascal} differential stress. The evolution of horizontal velocities is consistent with a very minor opening of axial mode I cracks, while the evolution of the sub-vertical velocity is explained by competing effects between the closure of sub-horizontal cracks under axial loading and the opening of axial cracks, as discussed above. When differential stress is increased above \SI{620}{\mega\pascal}, the slightly sharper decrease in all velocities ($\leq$1$\%$) is interpreted as the onset of strain localisation. This interpretation is mostly supported by the divergence and instability of local strain gauge data. The main observation from Figure \ref{fig:Vfault} is that, between the onset of strain localisation and up to the quasi-static failure of rock, velocities across and along the fault plane decrease more than the velocity out of the fault plane; however, both velocities across and along the fault plane decrease by the same amount (about 1$\%$). These combined observations reveal that ``mode I'' microcracking is not only very limited but also extremely localised at rock failure.
  
  \subsection{Micromechanics of Non-Dilatant Brittle Deformation in Antigorite}
  \label{subsec:discuss_mechanism}

  The brittle deformation in antigorite at room temperature and low confining pressure, and the mechanism responsible for it, appears to have several peculiarities that are unique relative to the behaviour of many polycrystalline rocks. These peculiarities are summarised in Table \ref{tab:combine} as revealed by combined observations and joint interpretations from stress-strain data, wave velocity measurements and microstructural observations during both direct and cyclic axial loading tests. The only way to reconcile the absence of volumetric dilation (even at low confining pressures) with the absence of stress-induced anisotropy and of notable stress-dependence of velocities prior to failure is that brittle deformation occurs purely by ``mode II'' shear microcracking, as proposed for antigorite by \citet{Escartinetal1997}. In addition, the propagation of shear microcracks occurs at very high stress and leads abruptedly to rock failure; it is not surprising that such highly unstable microcracking is then extremely localised. This interpretation is well supported by microstructural observations of a very localised damage zone formed of shear microcracks (see section \ref{subsec:res_micro}). Such ``mode II'' microcracks are only observed in antigorite specimens brought to failure, and not in specimens cyclically loaded to 90--95$\%$ of their mechanical strength.

  \begin{table*}
    \caption{Peculiarities of brittle deformation in antigorite (left column) and supporting evidence from mechanical (stress-strain) measurements, P- and S-wave velocity measurements, and microstructural observations on specimens recovered after deformation. References to relevant figures are given.}
    \label{tab:combine}
    \centering
    \begin{tabular}{|p{3cm}|p{3cm}p{3.5cm}p{3.5cm}|}
      \hline
      & Mechanical measurements & Wave velocity measurements & Microstructural observations\\
      \hline
      \emph{i)} \emph{Non-dilatant} inelastic behaviour and brittle deformation &
                                                                                  No dilatancy prior to rock failure (Figures \ref{fig:meca_direct}, \ref{fig:meca_cyclicII} and \ref{fig:meca_cyclicIV}), only inelastic axial strain (Figures \ref{fig:meca_cyclicII} and \ref{fig:meca_cyclicIV}) &
                                                                                                                                                                                                                                                                                                       No significant decrease in velocities and absence of stress-induced anisotropy prior to failure (Figures \ref{fig:velocP_direct}, \ref{fig:velocS_direct}, \ref{fig:veloc_cyclicII} and \ref{fig:veloc_cyclicIV})& No observation of ``mode I'' opening of microcracks (Figures \ref{fig:micro_deformed} and \ref{fig:micro_deformed_extra})\\
      \emph{ii)} \emph{``Elastic-brittle''} behaviour, with abrupt and unstable microcracking &
                                                                                                Yield point occurs very close to rock failure (Figure \ref{fig:meca_direct})&
                                                                                                                                                                              Small decrease in velocities occurs very close to rock failure (Figures \ref{fig:velocraw_direct}, \ref{fig:velocP_direct} and \ref{fig:velocS_direct})&
                                                                                                                                                                                                                                                                                                                                       No observation of microcracks after cyclic loading at 90--95$\%$ of rock strength\\
      \emph{iii)} \emph{Very localised} microcracking & No ``bulk'' dilatancy prior to failure (Figures \ref{fig:meca_direct}, \ref{fig:meca_cyclicII} and \ref{fig:meca_cyclicIV})&
                                                                                                                                                                                     No significant decrease in velocities even along fault plane (Figure \ref{fig:Vfault})&
                                                                                                                                                                                                                                                                             No observed damage except 100--200~\si{\micro\meter} near fault zone (Figures \ref{fig:micro_deformed} and \ref{fig:micro_deformed_extra})\\
      \emph{iv)} Brittle deformation accomodated by ``mode II'' \emph{shear microcracking}&
                                                                                            No dilatancy prior to rock failure (Figures \ref{fig:meca_direct}, \ref{fig:meca_cyclicII} and \ref{fig:meca_cyclicIV}), only inelastic axial strain (Figures \ref{fig:meca_cyclicII} and \ref{fig:meca_cyclicIV})&
                                                                                                                                                                                                                                                                                                                No significant decrease in velocities prior to failure (Figures \ref{fig:velocP_direct}, \ref{fig:velocS_direct}, \ref{fig:veloc_cyclicII} and \ref{fig:veloc_cyclicIV})&
                                                                                                                                                                                                                                                                                                                                                                                                                                                                                          Observation of shear microcracks in damage zone near localised fault (Figures \ref{fig:micro_deformed} and \ref{fig:micro_deformed_extra})\\
      \hline

    \end{tabular}
  \end{table*}

  Shear microcracks form almost exclusively parallel to (001) ``corrugated'' cleavage plane of the antigorite grains, in agreement with previous observations of \citet{Escartinetal1997} at room temperature and similar pressures. This occurs even when the (001) basal plane is not in a favourable orientation for sliding, such as at parallel or perpendicular to the compression axis (see \ref{subsec:res_micro}). Sliding on microcrack surfaces must necessarily be accomodated by compatible displacements or cracking within adjacent grains or at grain boundaries. A considerable number of experimental studies and micromechanical models have related the stress-induced microcracking and brittle behaviour of many polycrystalline rocks to the opening of ``mode I'' microcracks due to local tensile stress concentrations at the tip of the cracks favourably sliding under shear stress (for a review, see \citet{PatersonWong}). However, in antigorite, a striking peculiarity is that shear cracks do not nucleate any ``wings'', nor induce observable opening of adjacent cracks, even at low confining pressures. This suggests that crack propagation in antigorite must be highly controlled by anisotropy of fracture toughness or, more generally, by crystal anisotropy, and that ``mode II'' shear microcracking is the most favourable mechanism. Based on the observation of microstructures in damaged zones in antigorite (Figures \ref{fig:micro_deformed} and \ref{fig:micro_deformed_extra}), it is interpreted that the nucleation of shear microcracks is accomodated by a combination of three mechanisms: shear displacements and/or shear microcracking along cleavage planes in adjacent grains, ``tearing'' along grain boundaries and, in the few tens of microns from the fault, delamination along cleavage planes of antigorite crystals associated with grain size reduction.

  The mechanism of ``mode II'' shear microcracking is entirely consistent with a positive dependence of antigorite strength upon confining pressure, as increasing confining pressure results in an increase in the normal stress acting on (existing or incipient) crack surfaces, which provides frictional resistance against sliding. In addition, the relatively small size and random orientation of antigorite grains forms a ``locked'' microstructure, which necessitates substantial stresses for crack sliding and propagation to overcome and potentially explains the high strength and very abrupt failure of the material.

  \section{Conclusions}
  We have presented the results of a broad experimental study that combines new measurements of P- and S-wave velocities, at four different orientations to the compression axis, with measurements of axial and volumetric rock strain during brittle deformation of antigorite-rich ($>$95$\%$) serpentinite specimens. Such new measurements were taken during hydrostatic loading and subsequent axial loading in the 20--150~\si{\mega\pascal} confining pressure range and at room temperature. In addition to direct loading tests up to rock failure, cyclic loading tests were conducted to interrogate rock inelasticity and, for some tests, to achieve quasi-static or ``controlled'' rock failure. Measurements on antigorite specimens have been compared to an additional direct loading test on a ``reference'' crystalline rock, Westerly granite, at 100~\si{\mega\pascal} confining pressure.

  \emph{Mechanical measurements} are in line with previous findings by \citet{Escartinetal1997} on the same antigorite serpentinite. The mechanical strength of antigorite samples is high, comparable to that of crystalline rocks, and broadly consistent with Byerlee's rule below \SI{200}{\mega\pascal}. At all confining pressures, the mechanical behaviour is ``elastic-brittle'' with a yield point occuring very close to an abrupt rock failure. Up to rock failure, brittle deformation of antigorite is non-dilatant and only a small amount of inelastic axial strain can be created. \emph{Wave velocity measurements} reveal new peculiarities of brittle deformation in antigorite. Up to failure, the variations of P- and S-wave velocities with differential stress are very small in all directions with respect to the compression axis. As a result, the seismic signature of brittle deformation in antigorite is characterised by a spectacular absence of stress-induced anisotropy. Such behaviour is markedly different from the one observed in granite, which reveals a large decrease of velocity from about 50$\%$ of failure stress, and 10$\%$ stress-induced anisotropy at failure. \emph{Microstructural observations} demonstrate that brittle deformation is extremely localised on a fault. Strain localisation results from the propagation of shear microcracks that are only observed within 100-200~\si{\micro\metre} from a fault zone, and on specimens loaded to complete failure. In addition, the behaviour and observed microstructures in slighly anisotropic antigorite specimens show very strong similarities with that of the isotropic antigorite specimens.

  The P- and S-wave velocity measurements confirm the classification of antigorite serpentinites as a ``high-velocity'', ``low'' Poisson's ratio end member within the serpentine group of minerals. The dependence of the wave velocities upon hydrostatic pressure was inverted to obtain quantitative estimates of crack density and aspect ratio distribution, based on a differential effective medium model for an isotropic solid containing an exponential distribution of randomly oriented spheroidal cracks. The existing or ``zero-pressure'' population of open microcracks in antigorite is broadly comparable to that of Westerly granite. However, in antigorite, mechanical and velocity data as well as microstructural observations all indicate a striking absence of stress-induced opening of ``mode I'' microcracks. As proposed by \citet{Escartinetal1997}, at low pressures and room temperature, brittle deformation in antigorite occurs purely by ``mode II'' shear microcracking, exclusively following the antigorite basal or ``cleavage'' planes. These new experimental constraints on the seismic signature of polycrystalline antigorite deformation, and of the micromechanical mechanisms responsible for it, remain to be complemented by additional laboratory measurements at higher confining pressures and temperatures.

  \begin{figure}[b]
    \centering
    \setfigurenum{A.1}
    \includegraphics[width=0.425\textwidth]{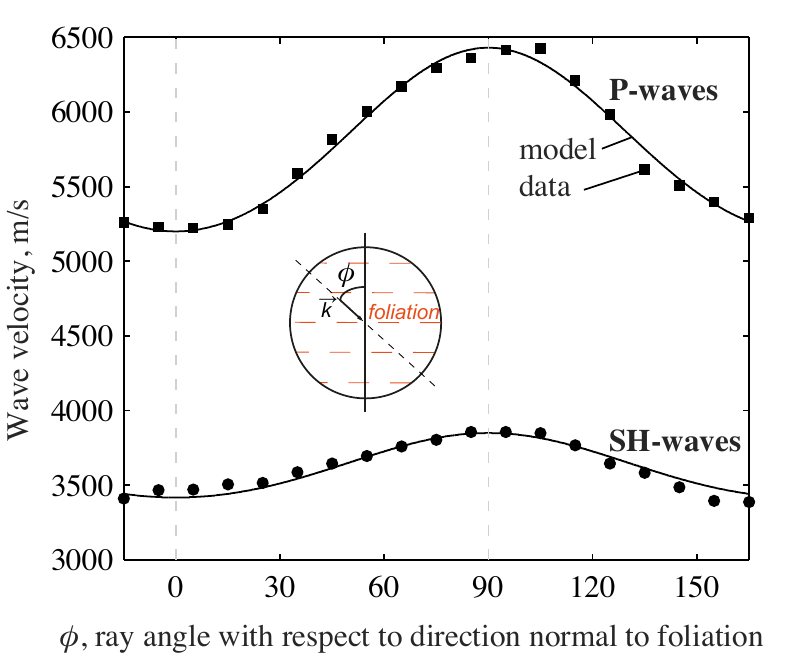}
    \caption{Angular variation of P and SH-wave velocities with respect to the direction normal to foliation (\ang{0}: perpendicular to foliation; \ang{90}: parallel to foliation), under ambient conditions. The velocity of P- and S-waves propagating along direction $\overrightarrow{k}$ was measured every \ang{10} around a cylindrical serpentinite specimen, cored from the ``anisotropic block'' in the out-of-plane direction, Figure \ref{fig:micro_asis}b. Squares and circles: experimental data; curves: best fit to data using model of elastic transverse isotropy (see section \ref{subsec:spec} for details). } 
    \label{fig:TI}
  \end{figure}
  \appendix
  \section{Improved Friction Corrections for Converting Externally Measured Load to Differential Stress on Rock}
  \label{sec:AppendixA}
  If $F_{\mathrm{meas}}$ denotes the load measured \emph{externally} by a load cell, the true load on the rock sample, $F_{\mathrm{samp}}$, is given by:
  \begin{equation}
    \label{eq:A1}
    F_{\mathrm{samp}}=F_{\mathrm{meas}}-\Delta F_{\mathrm{f}},
  \end{equation}
  where $\Delta F_{\mathrm{f}}$ denotes the total \emph{frictional resistance}  along the loading column. It is found that friction depends on both confining pressure and axial load as:
  \begin{equation}
    \label{eq:A2}
    \Delta F_{\mathrm{f}}=F_{\mathrm{f}}(P_{\mathrm{c}})+ b \Big[F_{\mathrm{meas}}-F_{\mathrm{f}}(P_{\mathrm{c}})\Big].
  \end{equation}
  The first correction term, $F_{\mathrm{f}}(P_{\mathrm{c}})$, is a constant during an experiment at a given confining pressure $P_{\mathrm{c}}$, and accounts for a slight balance imperfection in the axial piston autocompensation chamber, and dynamic friction between axial piston and O-ring seals. $F_{\mathrm{f}}(P_{\mathrm{c}})$ is easily measured in the early stage of an experiment, while the piston is advancing prior to the hit point between piston and rock sample. The second correction term is load-dependent, and reflects the non-negligible additional frictional resistance (along O-ring seals) due to Poisson's expansion of the loading column as load increases. Correction parameter $b$ is accessible by reversing the piston at different loads -- a situation occuring at the onset of unloading in cyclic loading experiments. When reversing piston direction, the measured force rapidly drops, until the rock sample begins to unload, by an amount equal to $2\Delta F_{\mathrm{f}}$. The advantage of using strain gauges on rock sample is that any change of stress on the rock can be locally monitored with high precision. The improved friction correction was robustly tested and constrained by ten cyclic loading experiments (see Table \ref{tab:samples}). Although the first term $F_{\mathrm{f}}(P_{\mathrm{c}})$ largely dominates the overall correction, it is found that the full correction given by equation \eqref{eq:A2} is a significant improvement considering the large loads involved in this series of experiments. Differential stress is simply obtained by dividing the corrected load by the sample cross-sectional area.



  \begin{acknowledgments}
    The UK Natural Environment Research Council supported this work through grants NE/K009656/1 to NB and NE/M016471/1 to NB and TMM. David Wallis greatly helped our microstructural investigations. Discussions with Greg Hirth considerably improved this manuscript. Steve Boon, John Bowles, Jim Davy and Neil Hughes (UCL) and Jonathan Wells (Oxford) provided technical support. Experimental data are available from the UK National Geoscience Data Centre (http://www.bgs.ac.uk/services/ngdc/accessions/index.html\#item110892) or upon request to the corresponding author.
  \end{acknowledgments}

    \begin{figure}[t]
    \centering
    \setfigurenum{A.2}
    \includegraphics[width=0.425\textwidth]{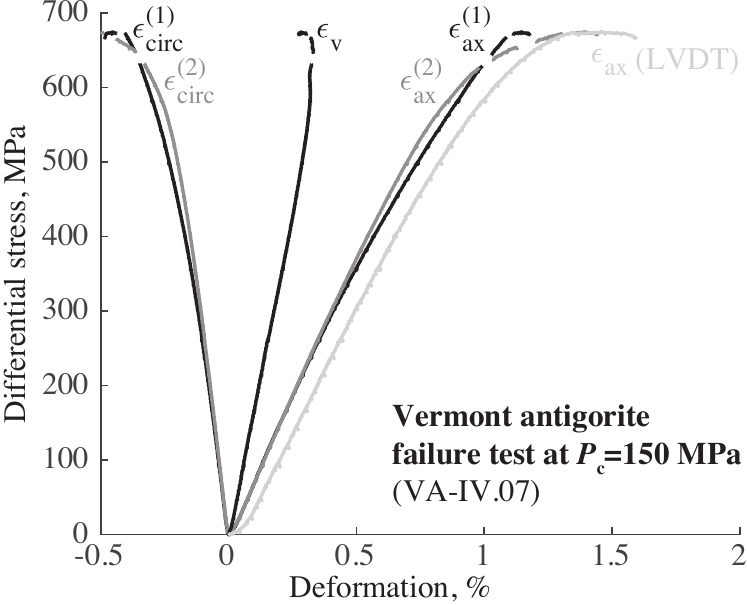}
    \caption{Stress-strain data for direct loading test at $P_{\mathrm{c}}$~=~\SI{150}{\mega\pascal}, until rock failure (specimen VA--IV.07). $\epsilon_{\textrm{ax}}$: axial strain; $\epsilon_{\textrm{circ}}$: circumferential strain. The superscript denotes a given pair of strain gauges. Dashed lines indicate final portion of the stress-strain curve where strain gauge data diverge from each other and from LVDT measurements, interpreted as strain localisation.} 
    \label{fig:mecaraw_direct}
  \end{figure}

\end{article}

\end{document}